\documentclass[traditabstract,twocolumn]{aa} 
\usepackage[normalem]{ulem}
\usepackage{graphicx}
\usepackage{txfonts}
\usepackage{subfigure}
\usepackage{longtable,lscape}
\usepackage{natbib}
\bibpunct{(}{)}{;}{a}{}{,} 
\begin{document}

   \title{A Normal Abundance of Faint Satellites \\
	in the Fossil Group NGC\,6482}


   \author{S. Lieder
          \inst{1,2}
	  \and S. Mieske\inst{2}
	  \and R. S\'{a}nchez-Janssen\inst{2,3}
	  \and M. Hilker\inst{4}
          \and T. Lisker\inst{1}
	  \and M. Tanaka\inst{5}
          }

   \institute{Astronomisches Rechen-Institut, Zentrum f\"ur Astronomie der
  Universit\"at Heidelberg, M\"onchhofstra\ss e 12-14, 69120
  Heidelberg, Germany, \email{slieder@ari.uni-heidelberg.de}
	      \and
	      European Southern Observatory, Av. Alonso de C\'ordova 3107, Vitacura, Santiago, Chile
	      \and
	      NRC Herzberg Institute of Astrophysics, 5071 West Saanich Road, Victoria, BC V9E 2E7, Canada
	      \and
	      European Southern Observatory, Karl-Schwarzschild-Strasse 2, 85748 Garching bei M\"unchen, Germany
	      \and National Astronomical Observatory of Japan, 2-21-1 Osawa, Mitaka, Tokyo 181-8588, Japan
             }

   \date{Received 14 February 2013; Accepted 3 September 2013}

 
  \abstract
  {Fossil groups are considered the end product in a galaxy group's evolution -- a massive central galaxy that dominates the luminosity budget of the group, as the outcome of efficient merging between intermediate-luminosity members. Little is however known about the faint satellite systems of fossil groups. Here we present a SUBARU/Suprime-Cam wide-field, deep imaging study in the $B-$ and $R-$band of the nearest fossil group NGC\,6482 ($M_{tot}\sim4\times10^{12}M_{\sun}$), covering the virial
radius out to 310\,kpc. We perform detailed completeness estimations and select group member candidates by a combination of automated object detection and visual inspection. A fiducial sample of 48 member candidates down to $M_R\sim-10.5\,$mag is detected, making this study the deepest of a fossil group up to now. We investigate the photometric scaling relations, the colour-magnitude relation, and the luminosity function of our galaxy sample. We find evidence of recent and ongoing merger events among bright group galaxies.
The colour-magnitude relation is comparable to that of nearby galaxy clusters, and exhibits significant scatter at the faintest luminosities. The completeness-corrected luminosity function is dominated by early-type dwarfs and is characterized by a faint end slope $\alpha=-1.32\pm0.05$. We conclude that the NGC\,6482 fossil group shows photometric properties consistent with those of regular galaxy clusters and groups, including a normal abundance of faint satellites.}


   \maketitle
%

\section{Introduction}
The study of so-called `fossil groups' (FGs) began about two decades ago. \cite{ponman_1994} found the first of these systems -- X-ray luminous galaxy groups characterized by their dominant bright central elliptical galaxy resulting in high mass-to-light ratios. The formal definition by \cite{jones_2003}, generally adopted by the community, is the following: (1) ensure that there is a dominant galaxy in the group by adopting an $R$-band magnitude gap $\Delta m_{12}$ of at least two magnitudes between the two most luminous galaxies. They restrict that criterion to galaxies within half of the group's projected virial radius $r_{vir}$ in order to ensure that $L^*$ galaxies have had enough time to merge due to dynamical friction -- since the merging timescale within $0.5r_{vir}$ for these systems is smaller than a Hubble time \citep{zabludoff_1998}. (2) Exclude `normal' elliptical galaxies that are not located in the center of the group by finding a hot gas halo that surrounds the galaxy (typical for central galaxies). This is achieved by a minimum X-ray luminosity of $L_{X,bol}=10^{42} h^{-2}_{50}$\,erg/s.\\
\cite{ponman_1994} interpreted their observations as witnessing the final stage in a group's evolution: an ancient stellar population in which most of the group's bright galaxies have merged into one luminous galaxy. Hence, they termed their finding a `fossil' group. This evolutionary scenario seems a plausible one for isolated parts in our Universe where galaxy groups can evolve undisturbed. Another FG formation scenario has been suggested by \citet{mulchaey_1999}. Here, FGs could merely be `failed' groups in which accidentally the majority of the baryonic mass was placed in a single dark matter halo - leading to the dominant central object. Investigations of FGs do not favor this scenario since several studies revealed that fossil brightest group galaxies (BGGs) exhibit disky isophotes \citep{khosroshahi_2006}. According to \citet{bender_1988} and \citet{khochfar_2005} it is considered that gas-rich mergers cause this isophotal behavior. This is also supported by \cite{aguerri_2011} and \cite{mendez-abreu_2012} who find that the S\'{e}rsic index of BGGs in FGs is significantly smaller than that in central galaxies of clusters. According to \cite{hopkins_2009}, small S\'{e}rsic $n$ are due to gas-rich mergers. \\
Investigations based on the criteria by \cite{jones_2003} generally led to the conclusion that FGs formed early and have not experienced any major merger event for several Gyrs \citep{sanderson_2003,khosroshahi_2004}. According to simulations they accreted the majority of their mass at high redshifts (e.g., $50\%$ at $z>1$; \citealp{donghia_2005,dariush_2007,diaz-gimenez_2008}). Fossil groups would hence constitute the top end of the hierarchical evolution of galaxies on group scales.\\
Some more recent studies suggest that FGs may only be a transient phase in a group's evolution \citep{von_benda-beckmann_2008,dariush_2010,cui_2011}. \citet{dariush_2010} argue that most of the early formed systems are not in a `fossil phase' at $z=0$, but indeed were at some earlier point during their evolution. It is also clear that the observational criteria for FG classification are to some extent arbitrary, and slight changes to, e.g., $\Delta m_{12}$ will change the fraction of environments classified as fossil. \cite{milosavljevic_2006} show that there is a smooth distribution of the luminosity gap among 730 SDSS clusters, in line with the idea that the observational definition of a fossil group does not necessarily highlight a marked change in underlying formation histories. Nevertheless, it is clear that fossil groups mark an extreme environment (in the tail of a smooth distribution), only expected in a few percent of massive dark matter haloes \citep{milosavljevic_2006}.\\
Here we are interested in the properties of the faint galaxy population in such an extreme environment. Early results suggested that FGs also lacked faint satellites \citep{jones_2000} providing potentially interesting constraints on the so-called `substructure crisis / missing satellite problem' of $\Lambda$CDM \citep{donghia_2004}. 
Recently, \citet{mendes_de_oliveira_2009} reanalyzed the FG used by \citet{donghia_2004} and found a steeper faint end slope of $-1.6$ -- comparable to clusters (e.g., Coma cluster: $\alpha=-1.4$, \citealt{secker_1997}). Other recent studies of FGs reveal more shallow faint end slopes of $\alpha=-1.2$ when determined out to $\sim r_{vir}$ \citep{cypriano_2006,proctor_2011,eigenthaler_2012}, whereas the restriction to $0.5r_{vir}$ even reveals declining faint ends, i.e., $\alpha>-1.0$ \citep{mendes_de_oliveira_2006,aguerri_2011,proctor_2011}.\\
However, none of these studies reaches magnitudes fainter than $M_R\sim-17$ mag\footnote{Deeper observations of FGs exist with the HST ACS, but their much more constrained spatial coverage makes them inappropriate for the study of widely distributed dwarf galaxies.}, barely scratching the dwarf regime. Therefore they cannot provide meaningful constraints on the asymptotic faint end slope of the galaxy luminosity function. In order to investigate this faint galaxy population of a FG we provide here a photometric analysis of the NGC\,6482 group down to $M_R\sim-10.5$ mag -- to our knowledge the deepest FG study yet.


\section{Data}
\subsection{NGC\,6482}
NGC\,6482 is the nearest known FG ($z=0.0131$; \citealt{smith_2000}) and therefore well suited for a ground-based study of its dwarf galaxy population. Adopting the cosmological parameters $H_0=70.0$ km\,s$^{-1}$\,Mpc$^{-1}$, $\Omega_{M}=0.3$, $\Omega_{\Lambda}=0.7$, NGC\,6482's distance is $m-M=33.7$ mag ($d=55.7$ Mpc) resulting in a physical scale of $0.263$ kpc\,arcsec$^{-1}$. We will use these numbers throughout the paper.\\
We anticipate here that we measure $M_R=-22.7\,$mag for the BGG and for the second ranked galaxy $M_R=-20.5\,$mag. Together with the X-ray luminosity of $L_{X}=1.0\cdot10^{42}$\,$h_{70}^{-2}$\,erg\,s$^{-1}$ \citep{bohringer_2000} NGC\,6482 meets the fossil definition by \citet{jones_2003}. Chandra observations imply a virial radius of 310 kpc, a hot gas mass fraction of $f_{gas}=0.16$ and a total mass of $M_{200}\approx4\times10^{12}M_{\sun}$, with an $R$-band mass-to-light ratio at $r_{vir}$ of $71\pm15$ $M_{\sun}/L_{\sun}$ \citep{khosroshahi_2004}. These data are consistent with the ROSAT study of \citet{sanderson_2003}, but we note that their larger field of view constraints results in a slightly larger $r_{vir}$ and in $f_{gas}\approx0.07$, more consistent with its $T_{X}\sim0.6$ keV.\\

\subsection{Observations}
On June 5th 2008 $R$-band images of NGC\,6482 were acquired using the Suprime-Cam wide field imaging instrument at the Subaru telescope (Table ~\ref{tab:timeline}). The Suprime-Cam camera is a mosaic of 10 2k$\times$4k CCDs with a pixel scale of 0.202 arcsec pix$^{-1}$ and covers an area of $34'\times27'$ per field \citep{miyazaki_2002}. The field-of-view corresponds to a physical scale of 624 x 458 kpc at the distance of NGC 6482, reaching the virial radius at the field edges. 85\% of the area which is enclosed by $r_{vir}$ is covered. After a chip replacement in July 2008, $B$-band images were obtained on August 4th 2008. Both observations were obtained in service mode under run ID S08B-150S (PI Hilker).\\
In either band, two short (R: 60s, B: 120s) and four long (R: 540s, B: 1020s) exposures were acquired, all centered on NGC\,6482. The average seeing in the $R$-band was 0.6 arcsec FWHM, and 1.0 arcsec in the $B$-band.\\

\subsection{Data reduction}

Suprime-Cam's reduction pipeline SDFRED was used to carry out overscan correction and flatfielding. As mentioned above, the $B$-band data were taken after a chip replacement. For this data set, each of the ten individual CCDs had four separate readouts, and thus four different gains. Unfortunately, accurate individual gains were not provided with the data\footnote{\emph{"For all data, S\_GAIN[1-4] and GAIN values at FITS header have errors larger than $10\%$. Those values are only for reference and should not be used for data analysis."} (http://smoka.nao.ac.jp/help/help\_SUPnewCCD.jsp)}. We adjusted the four gains within each CCD relative to each other such that sky brightness differences were less than $1\%$ after the flatfielding step. This adjustment was determined with the long exposures for which gain variations dominate the relative count difference between the four stripes per chip. The relative gain corrections were then applied to all other $B$-band exposures (flats, standards and short science exposures). The knowledge of the absolute gain value for the $B$-band data is not necessary because absolute photometric calibration was achieved with the standard star exposures, themselves being corrected with the same relative gain as the science data.\\
The THELI image reduction pipeline \citep{erben_2005} then was used for the remaining steps of data preprocessing. The astrometric calibration from the THELI reductions was based on cross-correlation with the PPMXL catalog of point sources using the scamp software. It also corrected for geometric distortions in the outermost parts of the Suprime-Cam fields. After the THELI photometry step, which is based on SExtractor \citep{bertin_1996} and adjusts the brightness levels of all chips and all exposures to each other, background subtraction was carried out using THELI.\\
After the THELI processing, instrumental magnitudes were computed from observations of standard stars taken in all four nights of the observing run, and the photometry calibrated on the Cousins $B$ and $R$ magnitude system of \citet{landolt_1992}.\\
The average 1$\sigma$ noise per pixel for the 36 minutes coadded image in $R$-band corresponds to a surface brightness of $\mu_R=27.2\,$mag/arcsec$^2$, and $\mu_B=28.4\,$mag/arcsec$^2$ for the 68 minutes composite image in $B$-band respectively. We thus have similar surface brightness sensitivities in $B$ and R, since typical early-type galaxy colours are around $B-R\lesssim1.5\,$mag.

\subsection{Calibration}\label{sec:calibration}
$R$-band and $B$-band data were observed in different nights. Each night, standard star fields were observed at two different points in time and for a range of different airmasses. At each airmass, two exposures per field were obtained. A significant extinction coefficient was found in the $B$-band, while for the $R$-band it was consistent with zero. Thus, the atmospheric extinction term in $R$-band is absorbed by the zero point. (Tab. \ref{tab:zeropoints}).\\
For the night June 5/6, 2008 when $R$-band data was taken, there were unfortunately some transparency variations due to the presence of clouds (Tab. \ref{tab:zeropoints}), which prompted us to take special care in the photometric calibration: the standard star exposures in the $R$-band around 03:00 Hawaii Standard Time (HST) showed huge sensitivity variations at an average zero point fainter than the mean of the other standard star exposures. We discarded these measurements and instead adopted the zero points measured two hours later around 05:00 HST, when conditions were stable. Fortunately the $R$-band science images, taken two hours earlier, were obtained at almost exactly the same airmass, so that we do not introduce a luminosity offset when not considering the atmospheric extinction term. Hence, the observations of multiple standards throughout the night allowed the exclusion of those standard images with a notable drop in throughput. The used calibration parameters are displayed in Tab. \ref{tab:zeropoints}. Systematic uncertainties due to photometric calibration arising from this table are given by $\sigma_R=0.04\,$mag and $\sigma_{B}=0.12\,$mag (uncertainties for Schlegel extinction are not provided).\\
We then proceeded to check for the presence of clouds in the $R$-band science data taken between 02:07 and 02:56 HST. In general, THELI compensates for varying atmospheric transmission during a sequence of images that are to be coadded. Relative flux offsets are determined and logged via the `fluxscale' parameter, from comparing the same sources in the individual images. All images are then normalized to the highest transmission within the stack of images. For a correct compensation of possible cloud effects, it is necessary to have at least one cloudless exposure in the stack of images which are coadded. THELI found that the first of the four long $R$-band exposures (540\,s) taken at HST 02:11 was indeed affected by a significant flux drop ($\sim 28\%$) with respect to the other three long exposures. This was corrected by THELI in the final coadded image via the `fluxscale' parameter. We consider this correction as robust, since the other long exposures had relative fluxes consistent with each other at the $5\%$ percent level.\\
The two short $R$-band exposures were taken immediately before the one long exposure that was apparently affected by reduced atmospheric transmission, between 02:07 and 02:11 HST. Among themselves, these short exposures (which will later be used only for fitting the centers of the brightest galaxies) did not show a notable relative variation in the flux. However, given their fast cadence, this does not exclude that they were affected by clouds. In order to test for presence of clouds, we run SExtractor on both the short coadded image and the long coadded image, both normalized to 1s integration time. The ratio of object fluxes between those two images shows that the short exposures had a 24\% lower sky transmission than the long exposures. When we used the short time exposures for our analysis -- only for the three brightest galaxies in our sample -- we corrected the $R$-band flux by that offset factor of 1.31 (=1/(1-0.24)).
For consistency reasons we applied the same procedure to the $B$-band data and found that the long time flux is $\sim 6\%$ less than the short time flux. This difference is within the typical variation in a clear night, suggesting that no clouds were present during these observations. To be consistent with the treatment in the $R$-band, we corrected all $B$-band longtime fluxes upwards by a factor of 1.06.
\begin{table}
\caption{$R-$band observation time line from 5 June 2008.}             
\label{tab:timeline}      
\centering                          
\begin{tabular}{c c c c c}        
\hline\hline                 
Image type & Exp. time & Airmass & HST & Note\\
\hline
Science & 60\,s & 1.024 & 2:07&\\
Science & 60\,s & 1.026 & 2:09&\\
Science & 540\,s & 1.036 & 2:11& (a)\\
Science & 540\,s & 1.048 & 2:21&\\
Science & 540\,s & 1.062 & 2:31&\\
Science & 540\,s & 1.079 & 2:47&\\
Standard & 5\,s & 1.258 & 2:56& (b)\\
Standard & 5\,s & 1.246 & 2:59& (b)\\
Standard & 5\,s & 1.064 & 4:47&\\
Standard & 5\,s & 1.064 & 4:48&\\
Standard & 2\,s & 1.063 & 4:50&\\
\hline                        
\end{tabular}\\
\tablefoot{HST: Hawaii Standard Time, (a): expected fluxscale deviates from THELI's applied fluxscale by $\sim28\%$, (b) large scatter in flux measurements of standard stars.}\\
\end{table}

\begin{table}
\caption{Photometric calibration parameters.}             
\label{tab:zeropoints}      
\centering                          
\begin{tabular}{c c c}        
\hline\hline                 
Filter & $R$ & $B$ \\
\hline
ZP [mag] & $27.37\pm0.02$ & $27.07\pm0.06$\\
$\kappa$ [mag] & --- & $-0.124\pm0.046$\\
X & $1.056\pm0.018$ & $2.006\pm0.318$\\
CT & $-0.001\pm0.019$ & $0.147\pm0.033$\\
A [mag] & $0.22\dots0.31$ & $0.35\dots0.50$\\
\hline                        
\end{tabular}\\
\tablefoot{ZP: zero point, $\kappa$: atmospheric extinction coefficient, X: mean airmass of exposures contributing to a coadded image, CT: colour term, A: Galactic extinction by \citet{schlegel_1998}} \\
\end{table}

\section{Sample selection and photometric procedures}
\subsection{SExtractor and morphological classification}\label{sec:sample}
\begin{figure}
  \includegraphics[width=0.5\textwidth]{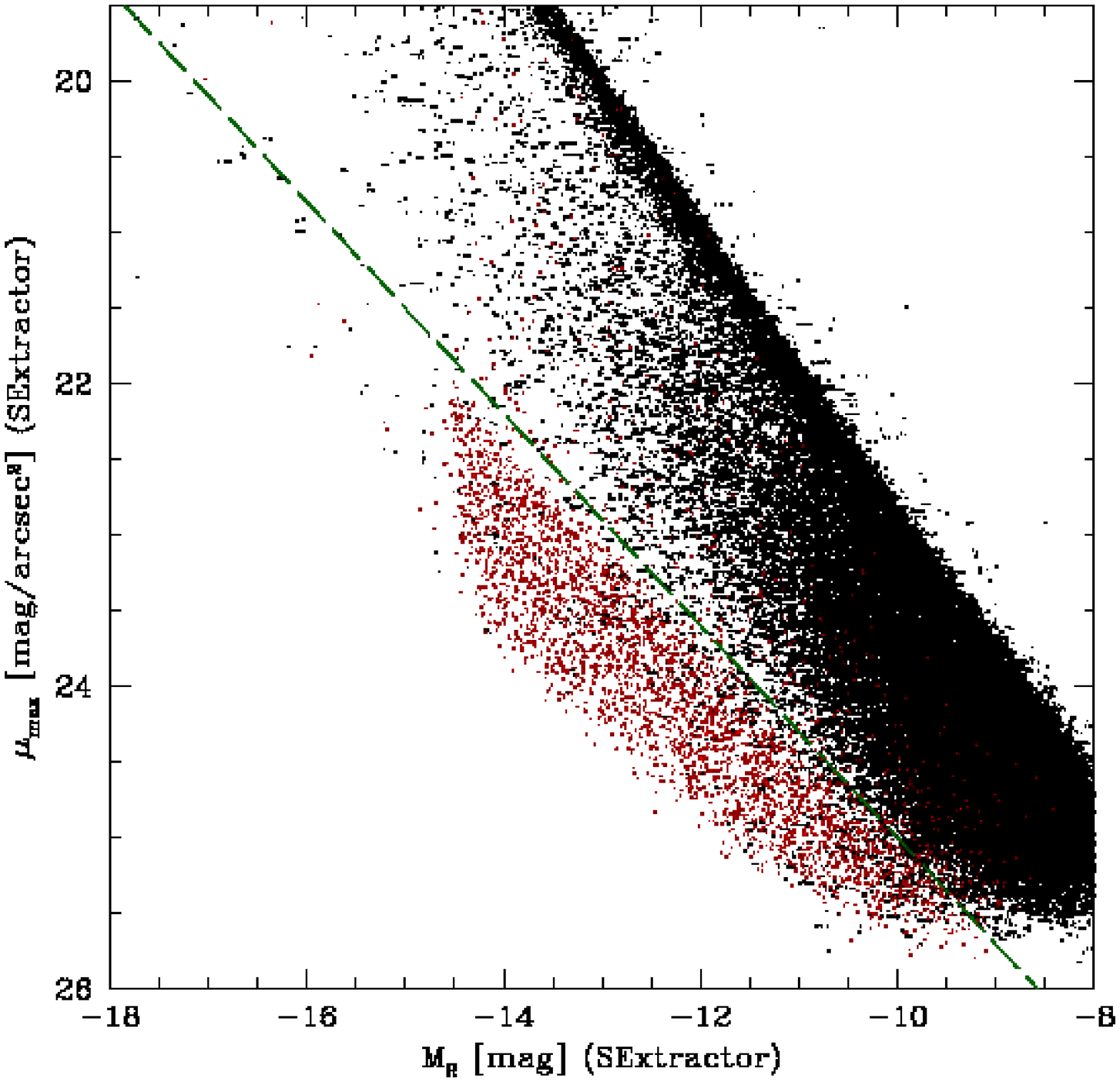}
  \caption{SExtractor MAG\_BEST and peak surface brightness are shown for all non-saturated objects detected in the long exposure of the Subaru NGC\,6482 field-of-view, including simulated dwarf galaxies with apparent sizes corresponding to the distance of NGC 6482. The SExtractor object detection parameters are given in Tab. \ref{tab:sextractor}. Black dots: all detected objects. Red dots: discovered simulated dwarf galaxies. Green dashed line: separation line adopted between galaxies and unresolved objects at the faint end $(\mu_{peak}=30+0.7\cdot M_R)$.}
   \label{fig:magmu}
\end{figure}
\begin{figure*}
   \subfigure[E]{\includegraphics[width=0.24\textwidth]{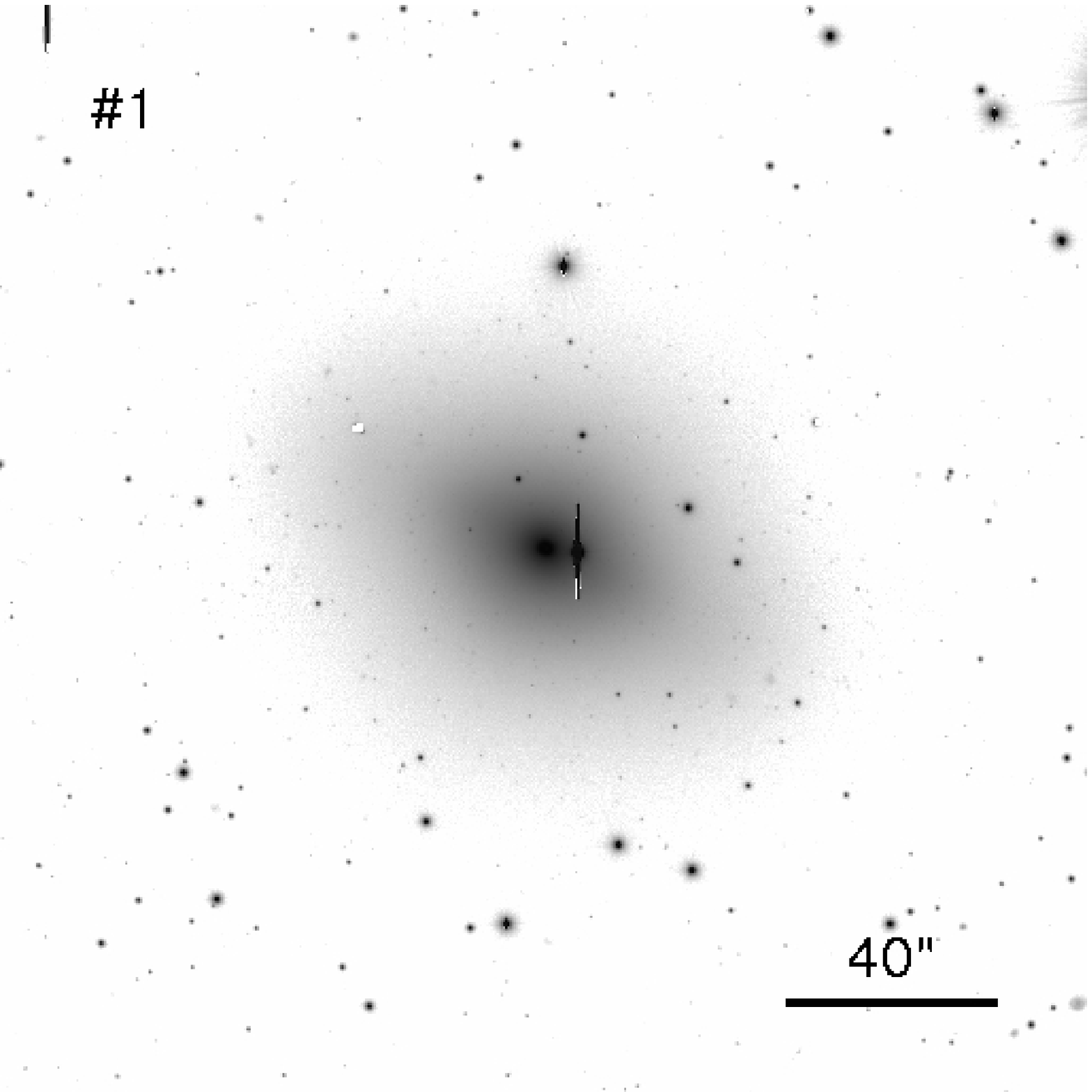}}\hfill
   \subfigure[SBc]{\includegraphics[width=0.24\textwidth]{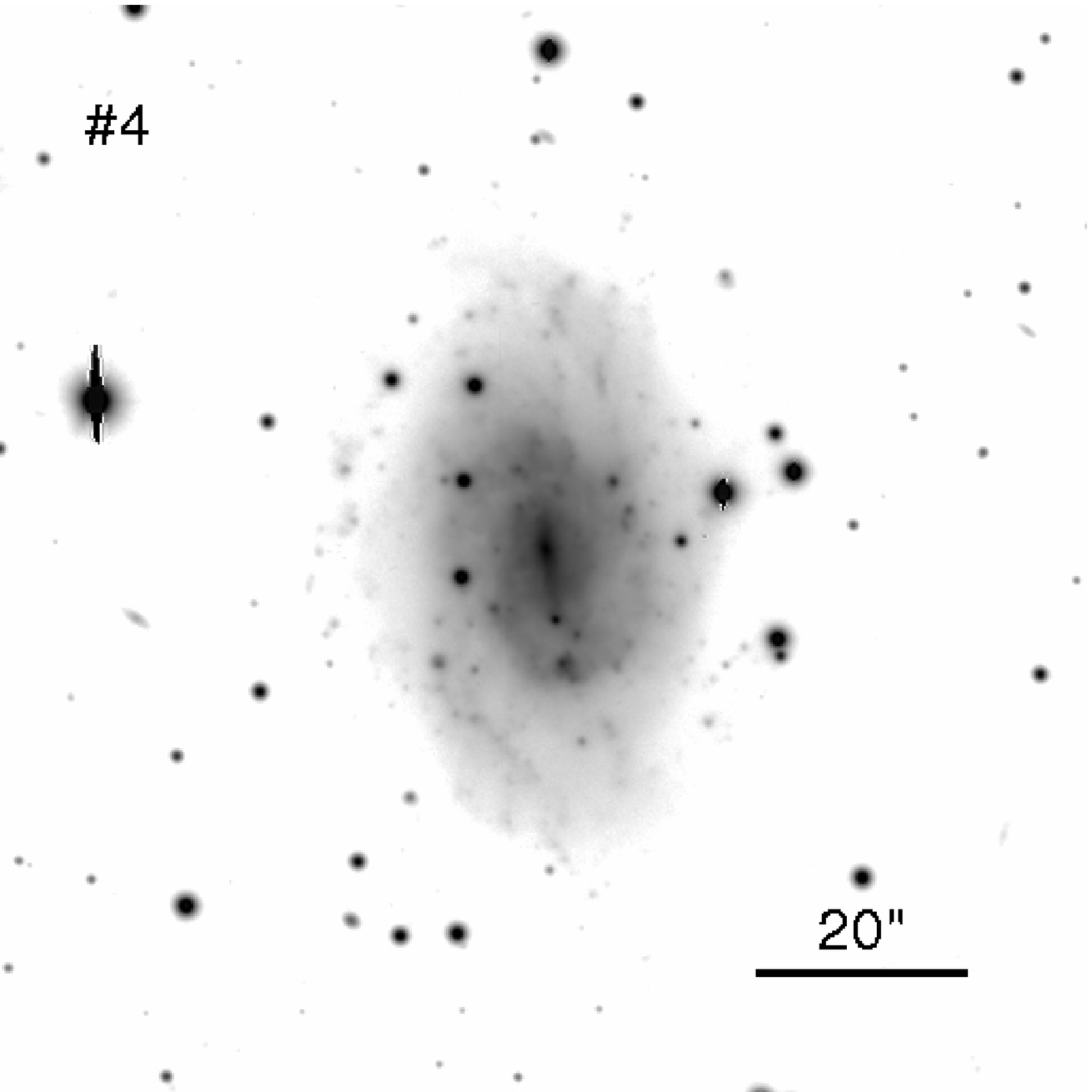}}\hfill
   \subfigure[dIrr]{\includegraphics[width=0.24\textwidth]{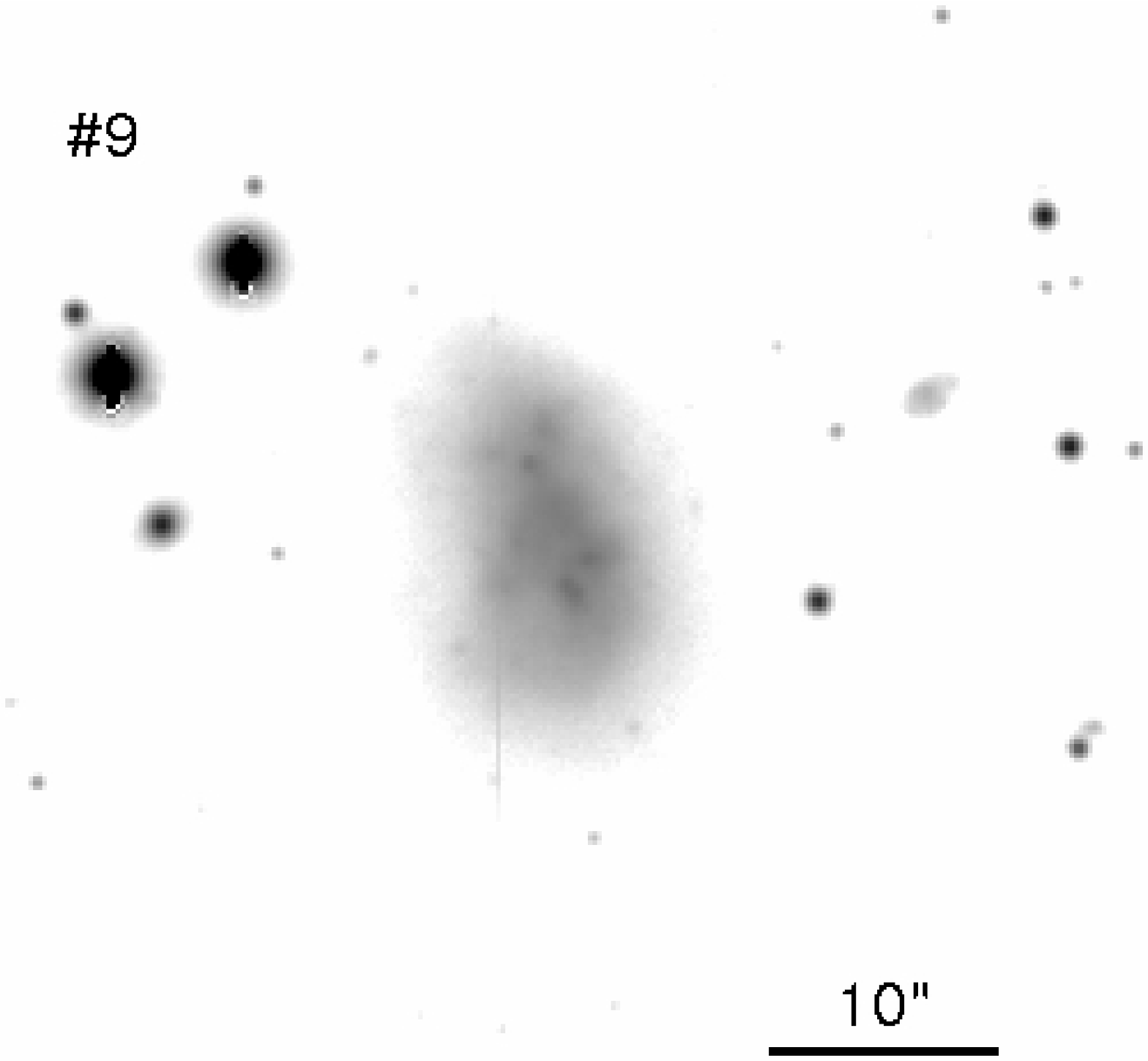}}\hfill
   \subfigure[dE,N]{\includegraphics[width=0.24\textwidth]{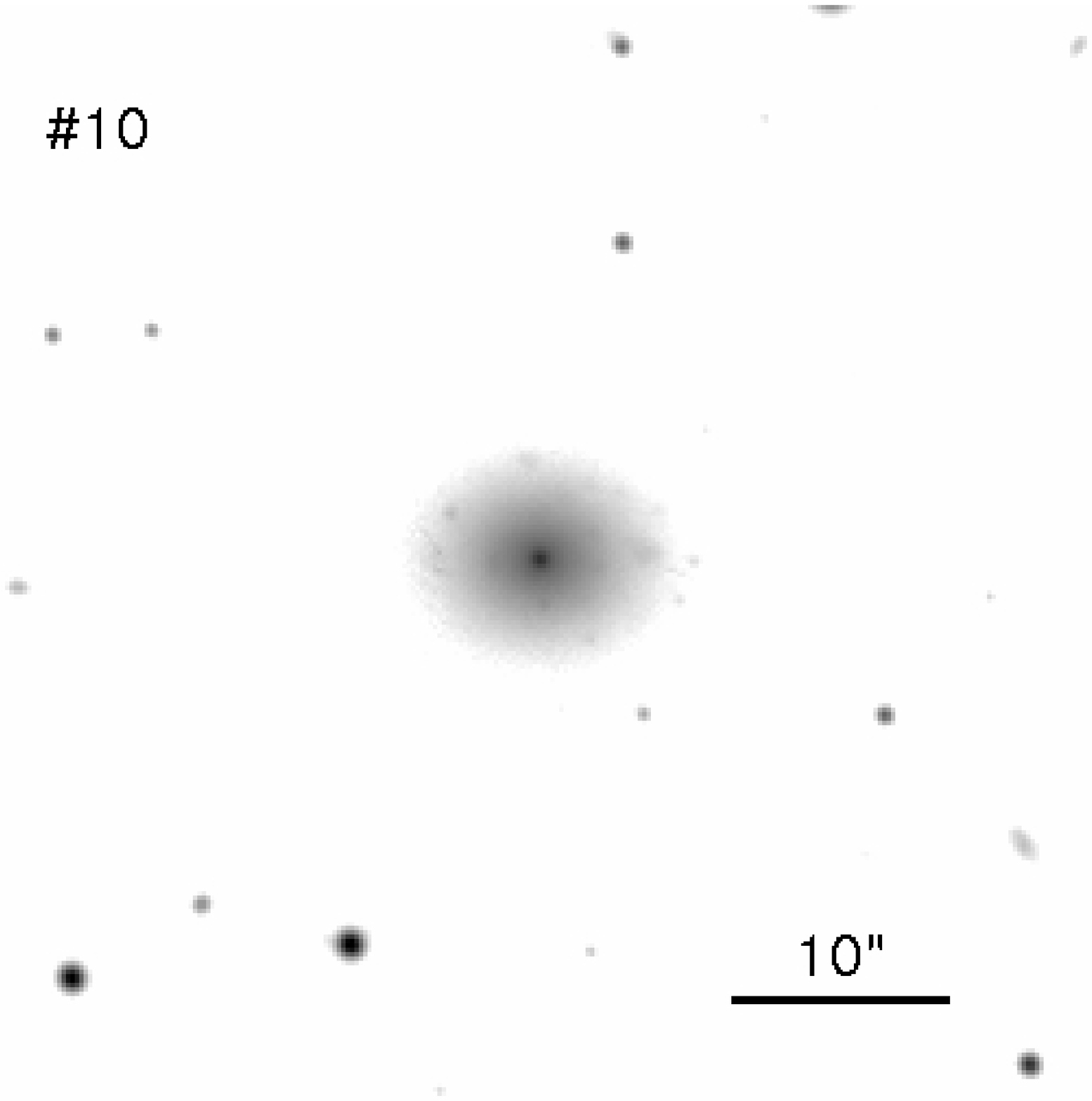}}\hfill
  \caption{Examples of morphological classification (provided in subcaption).}
    \label{fig:classification}
  \end{figure*}

SExtractor \citep{bertin_1996} was used for the detection of dwarf galaxy candidates in the field of view, followed by visual inspection. In order to optimize SExtractor's parameters we simulated seeing convolved dwarf galaxies with exponential surface brightness (SB) profiles and circular morphology (ellipticity=0) and added them to the $R$-band Subaru field. We put only 50 artificial galaxies on randomly chosen fields of 1 arcmin$^2$ to not saturate the already crowded field with objects. That process was repeated 100 times. Later in this section, we also use these simulations for completeness determination.\\
Dwarf galaxies are found to be relatively homogeneous in terms of their photometric scaling relations \citep{misgeld_2011} among different environments. As input for our simulations we used the $\mu$-mag relation found by \citet{misgeld_2009} for dwarfs in the Centaurus cluster\footnote{$\mu_{V,0}=0.57\cdot M_V+30.90$ and converted to $R$-band by adopting $V-R=0.6\,$mag} with a simulated scatter in $\mu$ of $\pm$ $0.96\,$mag around the fiducial relation. This scatter corresponds to the $2\,\sigma$ width of the relation found by \citet{misgeld_2009}. The goal of this exercise is to determine the location of typical dwarf galaxies in SExtractor's \verb*#MU_MAX-MAG_BEST# space, in order to establish a distinction from the crowd of small faint objects whose apparent sizes are close to the resolution limit of our data (Fig.\,\ref{fig:magmu}). The parameter space occupied by the simulated dwarf galaxies is clearly offset from the bulk of faint sources, which allows to define a fiducial separation line as indicated by the green dashed line in Fig.\,\ref{fig:magmu}. All objects below the green line in that plot are considered as possible members of the NGC\,6482 group. A small minority of simulated dwarf galaxies have recovered SExtractor parameters above that line, due to being superposed on another object -- mostly a brighter foreground star.\\
  \begin{figure}
    \centering
    \includegraphics[width=0.5\textwidth]{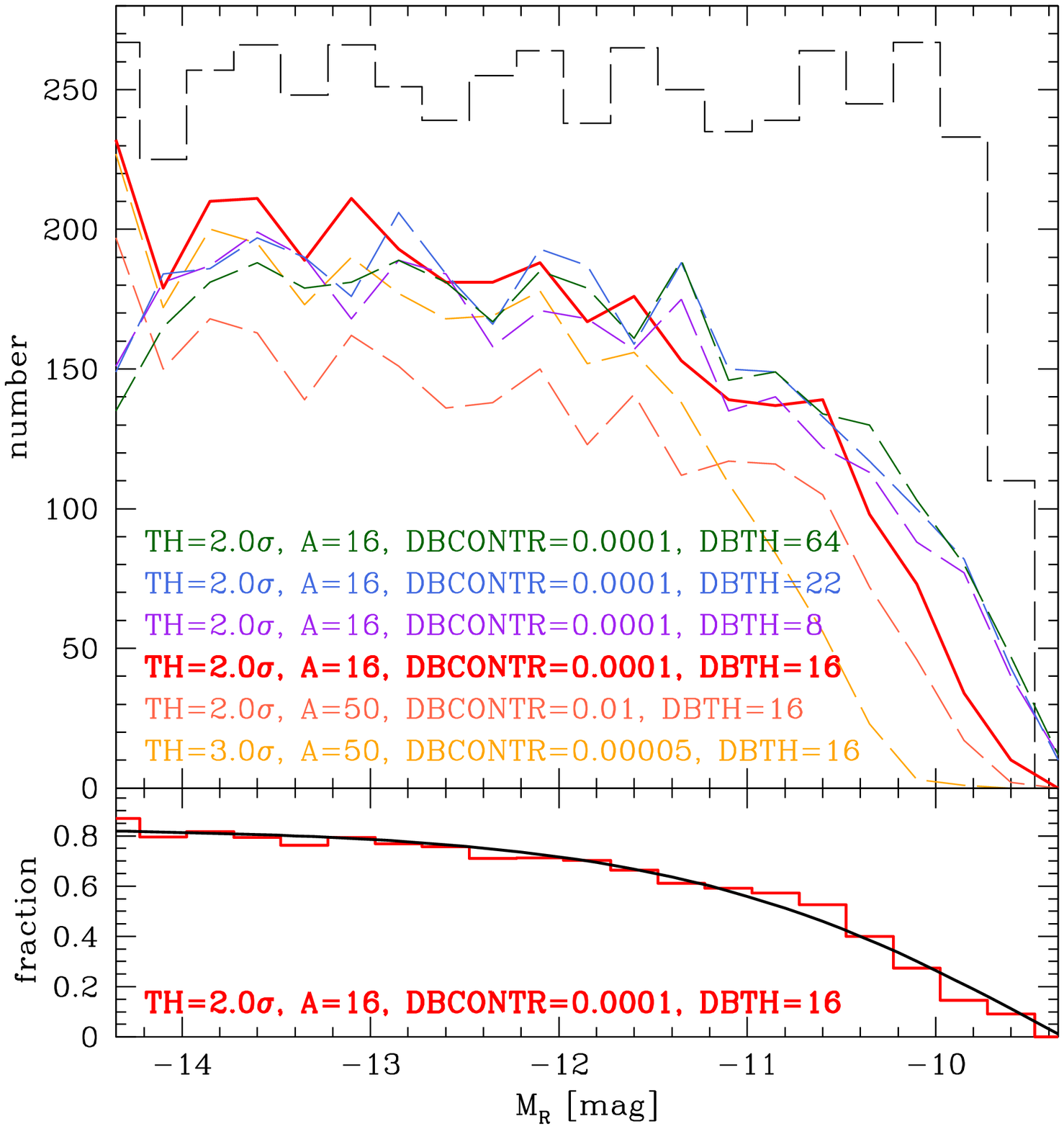}
    \caption{Determination of optimal SExtractor parameters. Upper panel: selection of tested SExtractor parameter sets. Coloured curves show SExtractor detections number counts for different setups (TH: detection threshold, A: minimum area of pixels above threshold, DBTH: number of deblending sub-threshold, DBCONTR: deblending contrast). The black, dashed histogram indicates the number of simulated galaxies in each magnitude bin. The red solid line represents our final choice, as it provides an optimal balance between the fraction of recovered objects, and the detection of spurious ones. Lower panel: the red, solid histogram shows the fraction of recovered simulated galaxies using the optimal parameter set. The black solid line is a fit to the red histogram (i.e., our completeness function). Due to crowding incompleteness (high foreground star density), the recovery completeness
saturates at 80\% for the brightest sources.}
    \label{fig:simulation}
  \end{figure}
  \begin{figure*}
    \centering
    \includegraphics[width=1.0\textwidth]{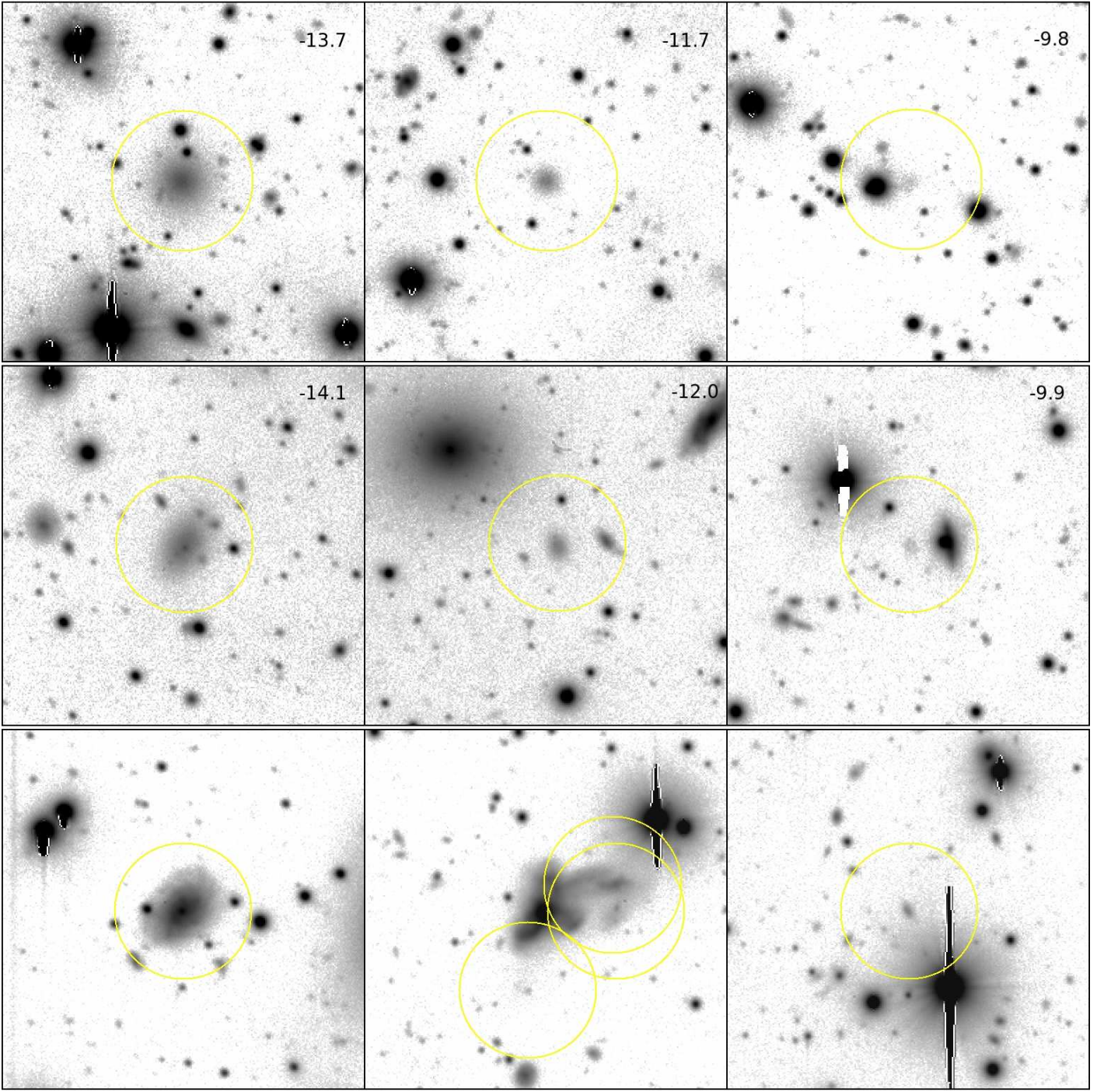}
    \caption{Illustration of example SExtractor detections. In each case the yellow circle has a diameter of 20 arcsecs and is centered on the object in question. Numbers in the upper right corner denote absolute $R$-band magnitudes. Upper panel: simulated galaxies placed into deep science images. Middle panel: objects accepted as possible members after visual inspection. Lower panel: rejected objects because of compact appearance or visible structure typical for background grand designed spirals or mergers.} 
    \label{fig:detections}
  \end{figure*}
Overall, the analysis of SExtractor's findings showed that even with extreme settings only up to $80-85\%$ of the simulated galaxies for the brightest galaxy bin are discovered (see red solid histogram in Fig.\,\ref{fig:simulation}). This is due to a crowding completeness limit imposed by the high foreground star density towards NGC\,6482 due to its low galactic ($l=48$, $b=23$) latitude (see e.g., Fig.\,\,\ref{fig:detections}). In the halo of bright stars, an automated detection algorithm like SExtractor tends to `overlook' individual sources. In addition to the crowding incompleteness, the usual surface-brightness incompleteness begins at $M_R\sim -13\,$ mag. We used an analytical expression to describe the surface brightness incompleteness, normalized to the crowding incompleteness level. This function is shown in Fig.\,\ref{fig:simulation} by the solid black line and reaches a $50\%$ completeness at $M_R\approx-10.5$\,mag and was used to correct the luminosity function (LF) for incompleteness.\\
The SExtractor detection parameters were varied (see Fig.\,\ref{fig:simulation} and Tab. \ref{tab:sextractor}) in order to find the optimal set in terms of recovered simulated galaxies with respect to all detections classified as galaxy -- given by the green line in Fig.\,\ref{fig:magmu}. We stress that for the actual photometry of dwarf galaxy candidates, individual aperture photometry is performed. The automatic magnitudes measurements by SExtractor are only used for the preselection of probable dwarf galaxy candidates.
\begin{table}
\caption{Optimized SExtractor parameters for dwarf galaxy detection.}             
\label{tab:sextractor}      
  \centering                          
  \begin{tabular}{l c}        
    \hline\hline                 
    parameter & value\\    
    \hline                        
    DETECT\_MINAREA & $16$ \\
    DETECT\_THRESH & $2.0\sigma$ \\
    DEBLEND\_NTHRESH & $ 16 $ \\
    DEBLEND\_MINCONT & $ 0.0001$ \\
    BACK\_SIZE & $12$\\
    BACK\_FILTERSIZE & $11$\\
    BACKPHOTO\_TYPE &  GLOBAL\\
    \hline                                   
  \end{tabular}
  \tablefoot{This parameter set is sensitive to faint extended objects but much more unresolved objects are found. Photometric output parameters are reliable in general, but individual outliers of up to $0.5\,$mag between input and recovered magnitude can occur due to the rather extreme setting of background determination.}
\end{table}
  \begin{figure*}
    \centering
    \includegraphics[width=1.0\textwidth]{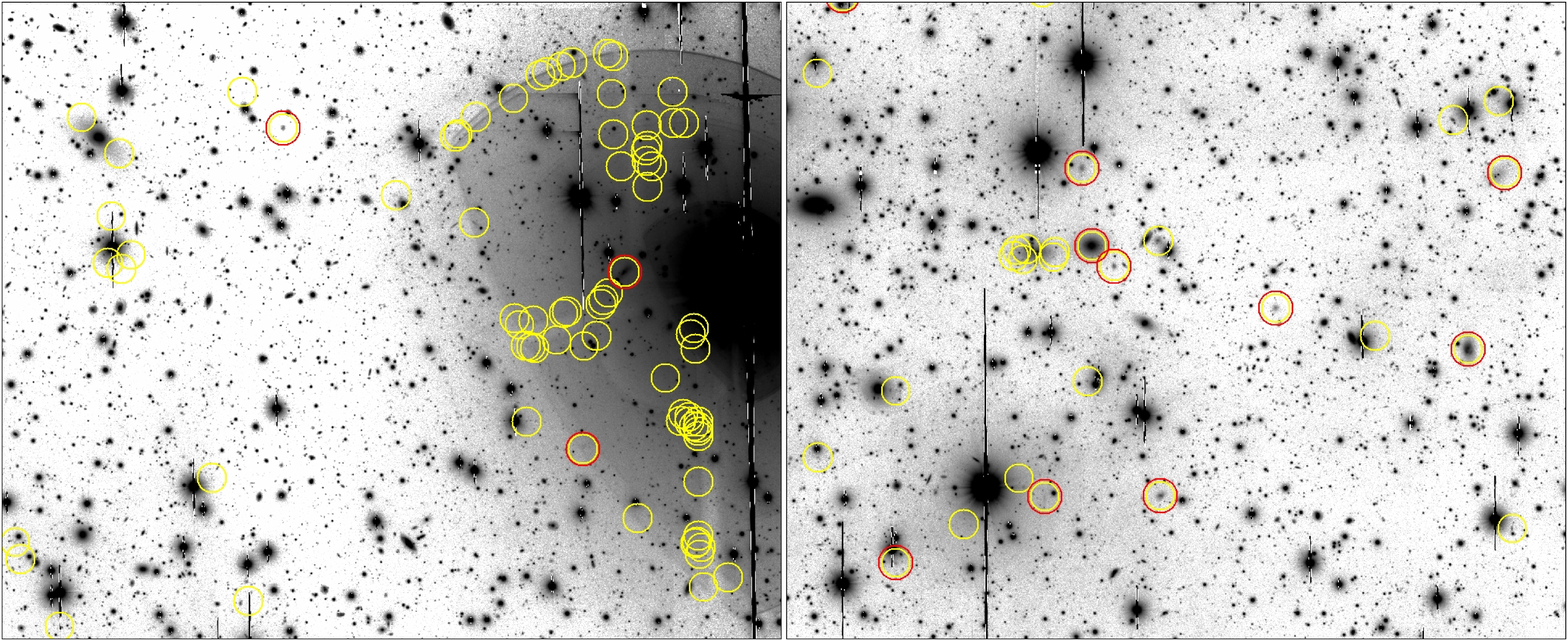}
    \caption{Illustration of preselected SExtractor detections considered as possible members (red circles), and artifacts/background sources (yellow circles). Left: plenty of objects are rejected since they are detected in the refraction halo of a foreground star and shadows along instrument suspension. Right: a field not affected by artifacts where detections close to spikes of foreground stars were rejected. Only a few objects remain and are considered as possible group members (red circles).}
    \label{fig:rejected}
  \end{figure*}
The optimal SExtractor parameter set for dwarf galaxy detection is displayed in Tab. \ref{tab:sextractor}, and yielded a total of 621 galaxies below the green line, out of a total of 61120 detections in the field of view. Those 621 objects are a preselection and were subsequently inspected visually by two of the authors (SL and SM) in an independent manner in order to reject artifacts (see Fig.\,\ref{fig:rejected}). The majority of sources were indeed readily identified artifacts in the halos of bright stars, see e.g., the left panel of Fig.\,\ref{fig:rejected}. Another goal of the visual inspection was the rejection of obvious background galaxies like low surface brightness spiral galaxies, interacting low surface brightness galaxies (merging events) or small very faint galaxies which appear too compact (compare right center panel and right bottom panel of Fig.\,\ref{fig:detections}). In particular the latter distinction has an impact on the number counts at the faint end of our sample. It is well known that faint dwarf galaxies (at $M_R\gtrsim -12\,$mag) are diffuse while their effective radius does not change significantly with luminosity (see \citealt{misgeld_2011}). Hence, we implicitly assume for our visual inspection that the faint dwarf galaxies should exhibit such a diffuse appearance.\\
Of the 621 SExtractor detections, 83 remained as visually confirmed candidate dwarfs. Twelve further sources were rejected because either a spike of a nearby foreground star covered the object's center, or structure was visible after the subtraction of the modeled galaxy (see Sect. \ref{sec:photometry}), or there was no or only a very weak ($\sim1\sigma$ above sky) counterpart found in $B$-band. Furthermore, two obvious dwarf galaxies (with $M_R\sim -11.5\,$mag and $-10.7\,$mag) not detected by SExtractor, and all bright galaxies (six galaxies brighter than $M_R=-18\,$mag and a diffuse dE with $M_R\sim-15.6\,$mag) were included to the sample since SExtractor was tuned to find faint galaxies with $M_R>-14\,$mag. Overall, 80 galaxies in the absolute magnitude range $-8.8\,$mag to $-22.7\,$mag at NGC 6482's distance were selected for detailed photometric analysis.\\
The morphological classification we adopted during visual inspection follows the extended Hubble scheme by \cite{sandage_1984}. In the dwarf regime we simplify it by labeling early-type dwarf galaxies generally "dE" and irregular dwarf galaxies "dIrr". Examples of our classification are shown in Fig.\,\ref{fig:classification}. The morphological classification type of each group galaxy is provided in Tab.\,\ref{tab:appendix2}.

\subsection{Colour magnitude diagram}
The morphologically pre-selected sample was cleaned of further probable non-members via colour selection criteria. In the following subsection \ref{sec:photometry} we describe the photometric procedures applied for the colour measurement, and in \ref{sec:cmd} discuss the selection of the fiducial sample based on the distribution in colour-magnitude space.

\subsubsection{Photometric procedure for magnitude and colour measurement}\label{sec:photometry}
  \begin{figure*}
   \subfigure{\includegraphics[width=0.33\textwidth]{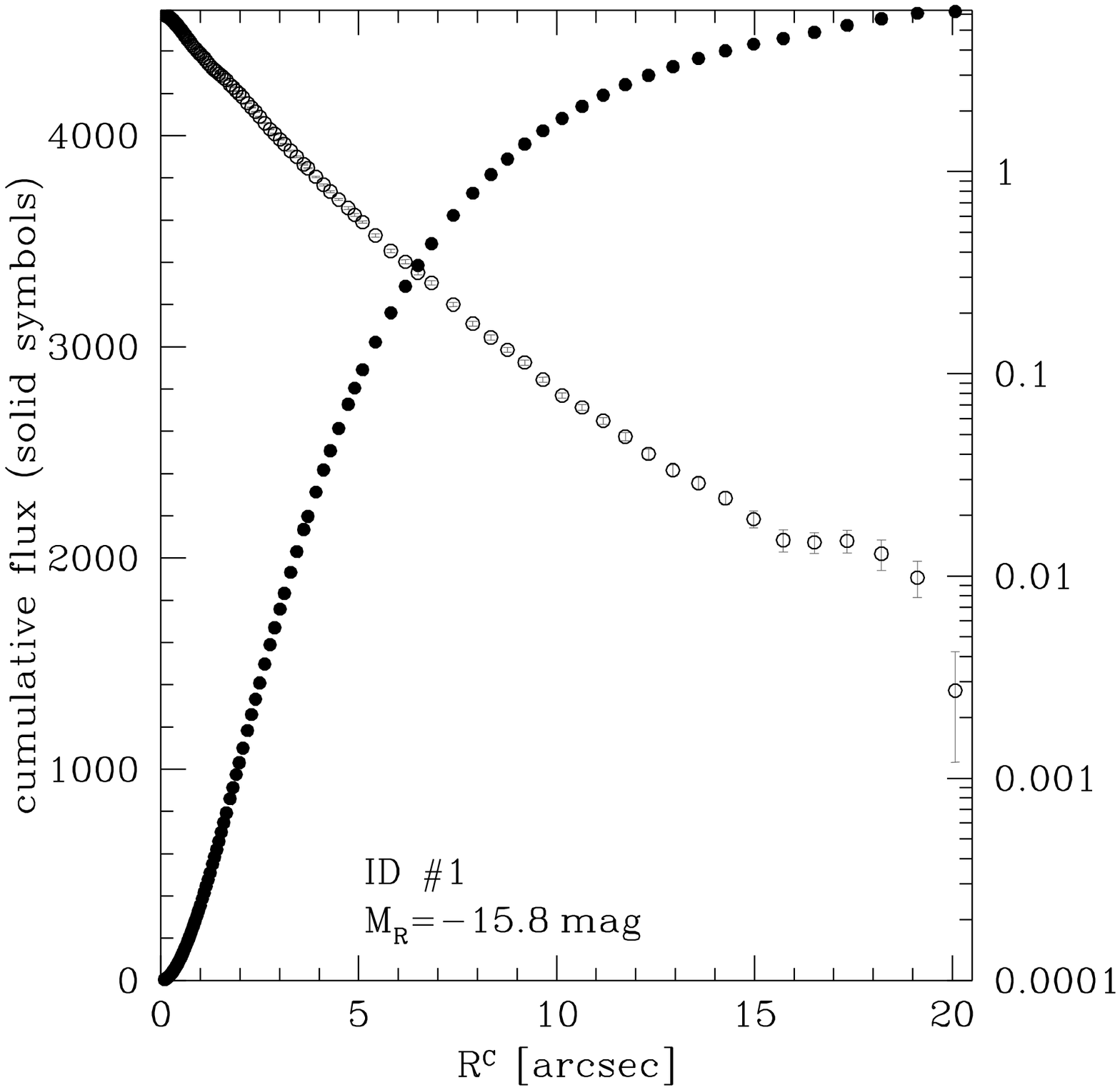}}\hfill
   \subfigure{\includegraphics[width=0.33\textwidth]{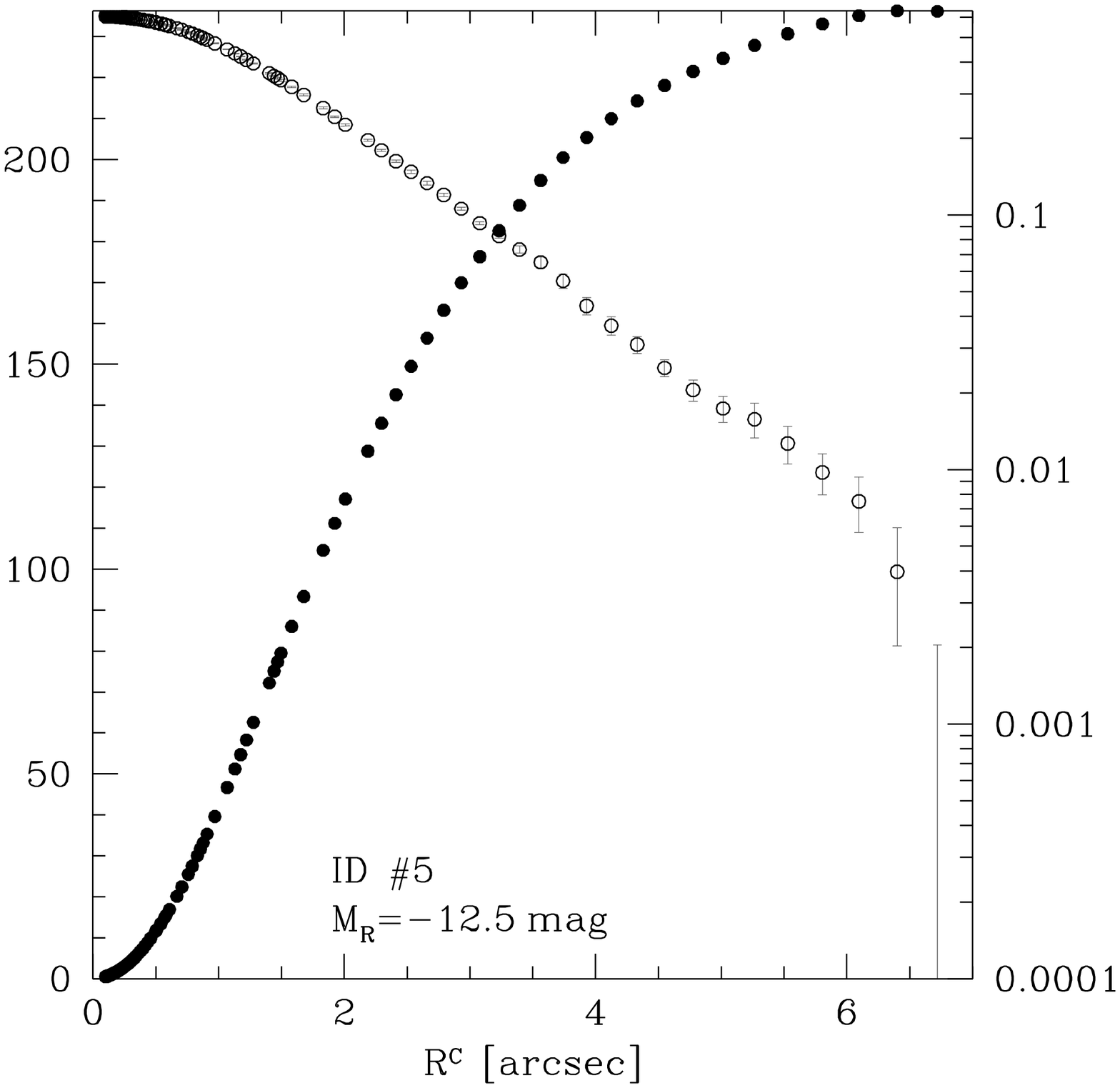}}\hfill
   \subfigure{\includegraphics[width=0.33\textwidth]{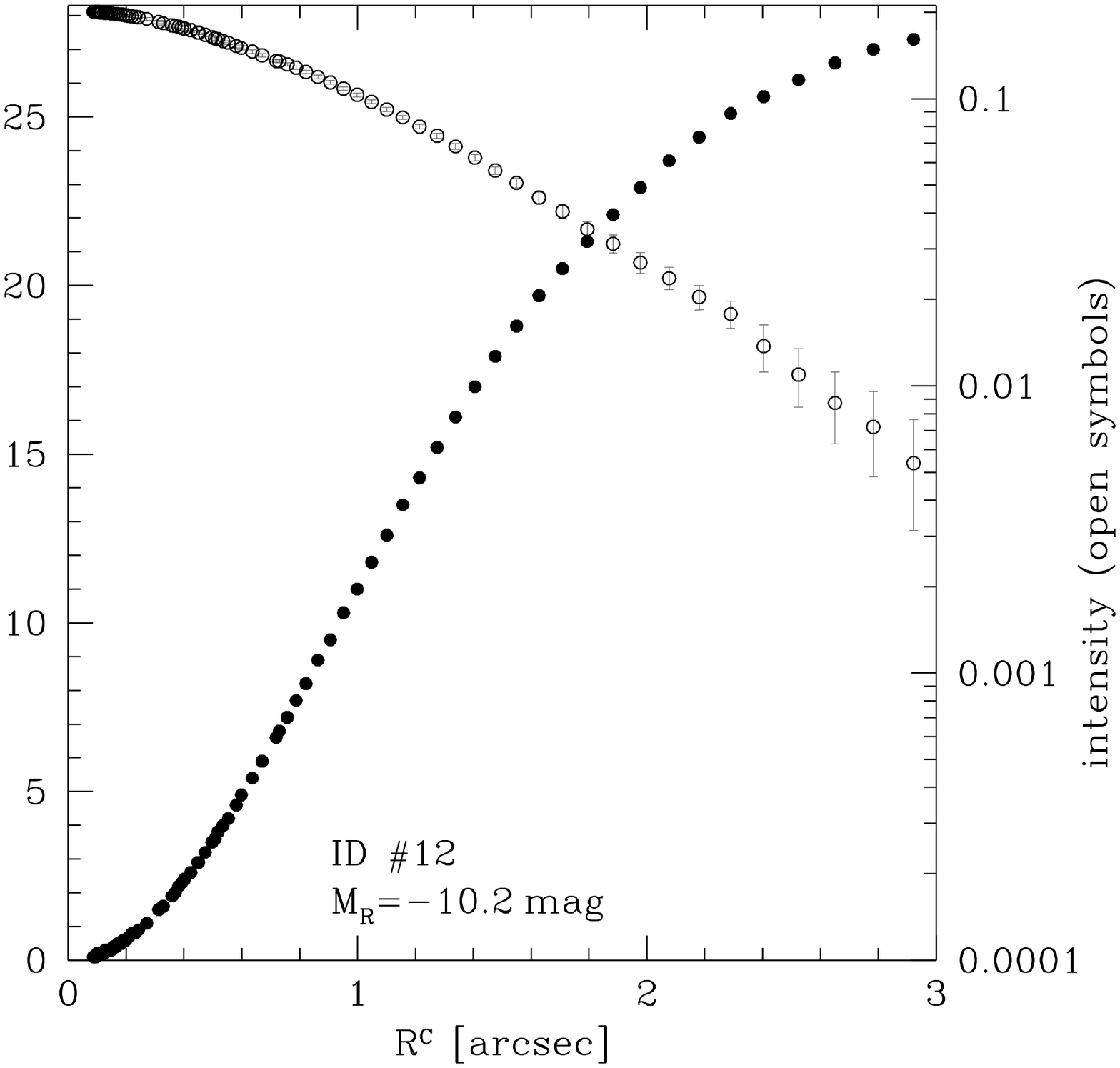}}\hfill
   \caption{Intensity profiles (open symbols, right y-axis) and cumulative flux (solid symbols, left y-axis) of three arbitrary chosen galaxies in the sample. Shown are all data points used for the determination of the total flux.}
   \label{fig:cog}
  \end{figure*}

  In order to  correct for the light blurring due to the PSF we degraded the $R$-band images to the worse seeing of the $B$-band images (1.0 arcseconds) using \verb*#IRAF# task \verb*#psfmatch#. Doing this we expect to measure the flux within the same physical isophotes. Since the observed field is very crowded, the light of a very bright foreground star reflection halo partially contaminates the light of a selected dwarf galaxy in many cases (see e.g., left image of Fig.\,\ref{fig:rejected}). In those cases the star was modeled in both passbands (using \verb*#ellipse#) and subtracted from the image in order to obtain more reliable values for the galaxy's luminosity. All galaxies were photometrically analyzed using the \verb*#ellipse# task \citep{jedrzejewski_1987} which is included in the \verb*#STSDAS# package of \verb*#IRAF#. All \verb*#ellipse# fits were performed with fixed parameters for center coordinates and position angle but variable ellipticity. In some cases, like the central galaxy NGC\,6482 itself, optimal results were obtained when the position angle was allowed to vary. Because the seeing of the $R$-band images was better, we did a first \verb*#ellipse# run for the undegraded $R$-band images. Those were used for the surface brightness analysis in Sec.\,\ref{sec:sb}. Another \verb*#ellipse# run was performed to fit the degraded $R$-band images. We then applied the obtained isophote table to the $B$-band image of the same object in order to measure the flux within the same physical isophotes. Obvious faint foreground stars and background galaxies were masked. \verb*#ellipse# fits were performed far beyond the galaxy's edge in order to see the signal reaches the amplitude of the background noise. Using that level, an individual background adjustment was done for every galaxy. The result of this background estimation is shown in the three panels of Fig.\,\ref{fig:cog}. The intensity (open circles) levels out at zero and associated error bars become as large as the signal, i.e., the signal is dominated by background noise (note the logarithmic scale of the intensity -- right axis).\\
   The \verb*#ellipse# output tables were used to determine all astronomical quantities which are presented in this study. The truncation radius of a galaxy was defined to be the last isophote at which the intensity is still larger than its error. The radial profiles of the cases shown in Fig.\,\ref{fig:cog} are displayed out to that truncation radius. The flux enclosed by that ellipse is used as (measured) total flux $f$. Using that flux the apparent magnitude of an object is calculated. Finally, the apparent magnitude of an object was corrected for galactic extinction, applying the foreground extinction map of \citet{schlegel_1998}. In Tab. \ref{tab:zeropoints} quantities determined for the photometric calibration are listed. The photometrical uncertainties of the dwarf spheroidals (dSph) are dominated by sky noise.\\
The half-light radius $r_{50}$ is determined as the radius enclosing 50\% of the measured total flux $f$. We determine the mean $B-R$ colour of each object using the flux within $r_{50}$ as determined in $R$-band. Within $r_{50}$ the signal-to-noise ratio is higher than that for the total flux. In light of the crowded field and the transparency variations (see Sect. \ref{sec:calibration}) in some cases a variable sky background can occur making $r_{50}$ more reliable for our colour estimation, especially for very low surface brightness targets for which total fluxes are difficult to determine.

\subsubsection{Fiducial sample definition via colour-magnitude selection}\label{sec:cmd}
  \begin{figure}
    \centering
    \includegraphics[width=0.5\textwidth]{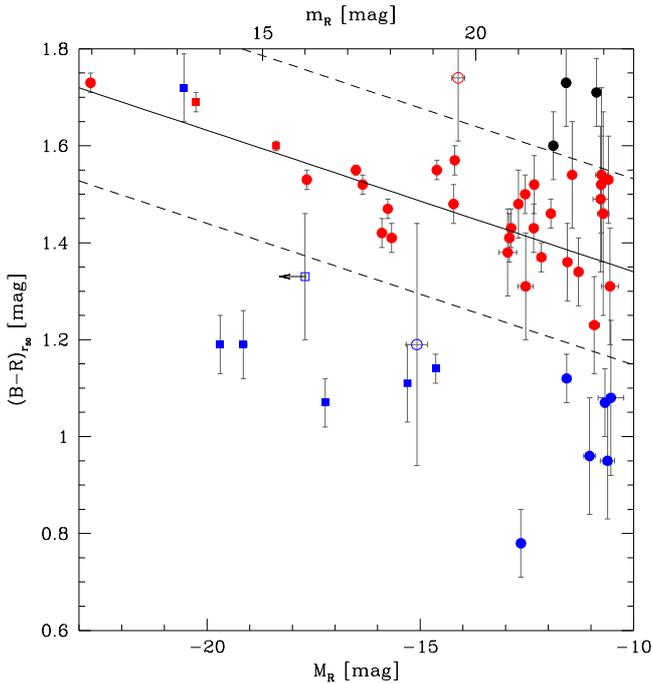}
    \caption{Colour-Magnitude Diagram of all objects brighter than $M_R=-10.5\,$mag (our $50\%$ completeness limit) and bluer than $B-R=1.75\,$mag. $B-R$ colours represent the integrated value within the half-light radius determined in $R$-band, except for the BGG --where the color is determined \emph{at} $r_{50}$--, and the disrupted galaxy (see Sect. \ref{sec:disrupted}). Red data points denote red sequence galaxies which we consider as group members (circles: elliptical galaxies, squares: S0 galaxies), blue data points for blue cloud galaxies (circles: blue dSphs, squares: spirals and irregulars), respectively. Black data points are galaxies considered to be background galaxies. The open blue square denotes the disrupted galaxy of Sect. \ref{sec:disrupted}. Open circles (one blue, one red) represent galaxies with uncertain photometry because the galaxy is superposed with a brighter object (BGG or refraction halo of a star) which could not be fully modeled. Because of their morphology those galaxies are included in the sample. The solid black line is a best fit for our red sequence CMR for galaxies brighter than $M_R=-14\,$mag (see text). The dashed lines illustrate the $3\sigma$ level of confidence which is used to reject background galaxies from the sample on the red side.}
    \label{fig:cmd}
  \end{figure}
The colours of all objects with $B-R<1.75\,$mag and brighter than $M_R=-10.5\,$mag are shown in the colour magnitude diagram (CMD) in Fig.\,\ref{fig:cmd}. We consider redder galaxies to be background contamination\footnote{A 12 Gyr-old stellar population with super-solar metallicity ($[Fe/H]=+0.2\,$dex) -- typical of luminous early-type galaxies -- has a $B-R$ colour of $\sim1.75\,$mag \citep{Worthey_1994}.}Galaxies fainter than this limiting magnitude have very large $B-R$ colour errors and are in the luminosity regime where the detection completeness is below 50\%. The colours represent the integrated value within the half-light radius $r_{50}$. Only the colour of the BGG, whose inner arcsec is saturated in $R$-band, is represented by its value at $r_{50}$. The $B-R$ value of the disrupted galaxy (at $M_R\approx-17.7\,$mag) comes from SExtractor analysis. It is clearly visible in the plot that three disky galaxies fall almost exactly on the red sequence (RS hereafter). These are the S0 host of the disrupted galaxy, an S(lens)0 and a dusty edge-on spiral -- the bright blue square in Fig.\,\ref{fig:cmd}. See ID 3, 6 and 2 in Tab.\,\ref{tab:appendix2}. The sample also contains two blue almost face-on spiral and three dIrrs denoted by filled blue squares in the CMD. The four brightest galaxies are spectroscopically confirmed group members (and indirectly the disrupted galaxy, see Sect.\,\ref{sec:disrupted}). Nicely visible in the CMD is the 2-magnitude-gap in $R$-band between the BGG and the second ranked galaxy -- one of the fossil criteria. Also noticeable is that the second brightest early-type galaxy has $M_R=-17.7\,$mag -- five magnitudes fainter than the BGG, already entering the dwarf galaxy regime.\\
To improve on the purely morphological constraints of group membership described in the previous section, we colour-restrict our sample to galaxies that lie within 3$\sigma$ of the red sequence (RS), whose location we determine with a least square fit to the data points shown in Fig.~\ref{fig:cmd}. In order to not be affected by the larger scatter of faint galaxies we fit the RS only to galaxies brighter than $M_R=-14.0\,$mag, and exclude the obvious blue sequence objects as well as three photometrically uncertain galaxies.\\
  \begin{figure}
    \centering
   \subfigure[dE(,N?)]{\includegraphics[width=0.24\textwidth]{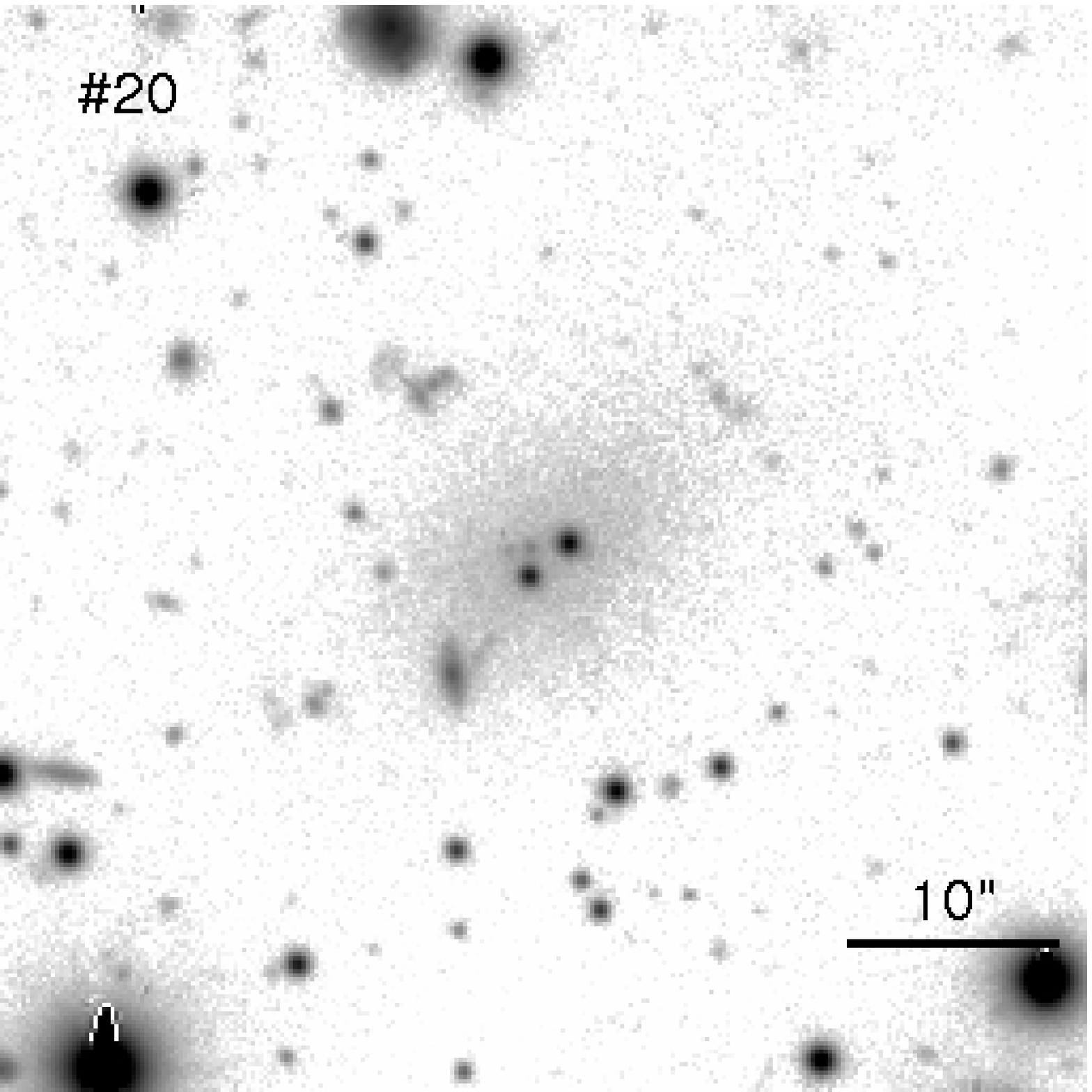}}\hfill
   \subfigure[dE,N]{\includegraphics[width=0.24\textwidth]{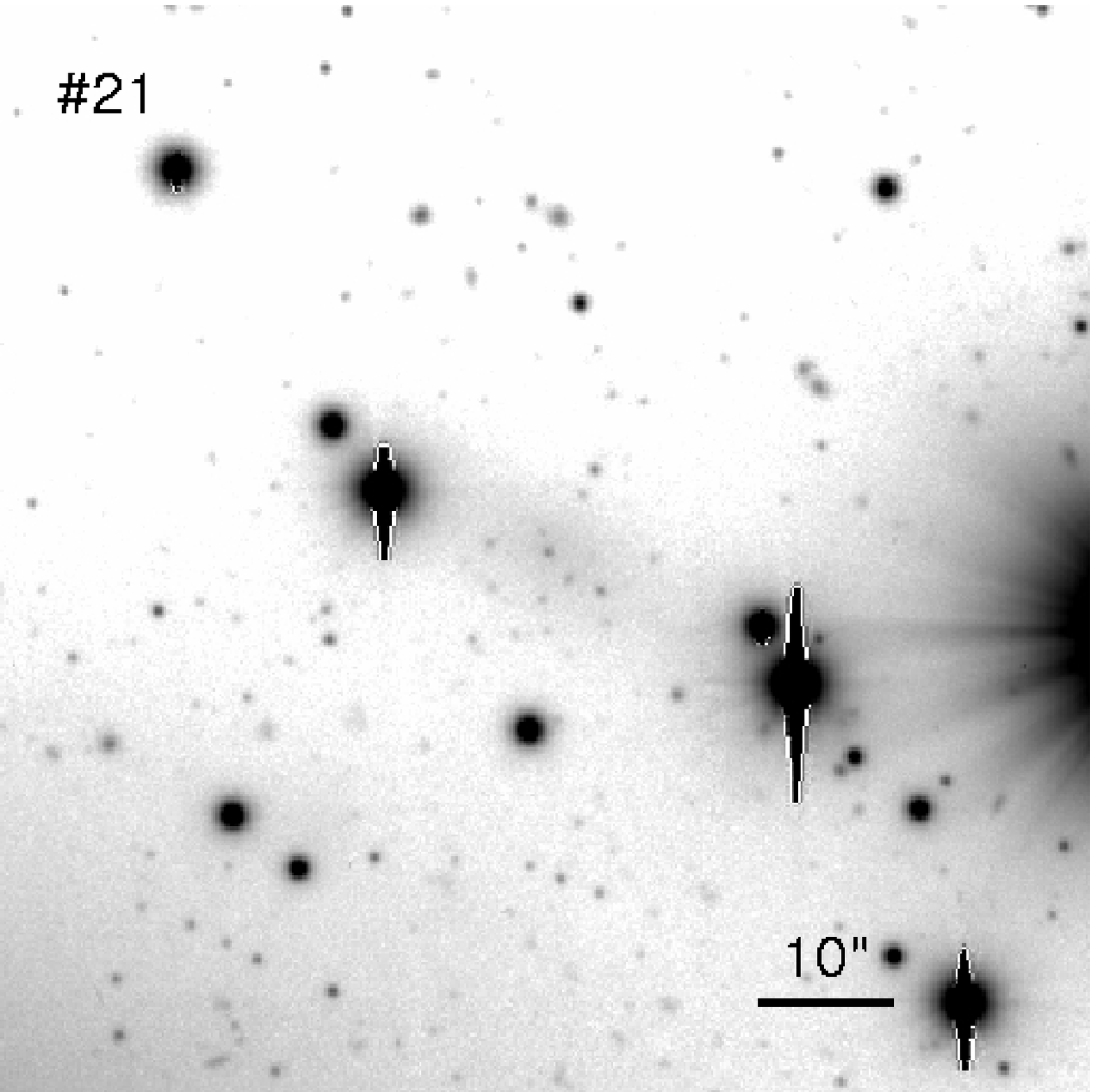}}\hfill
        \caption{Galaxies with uncertain photometry due to BGG's halo and several stars in close proximity. We believe both galaxies are part of the group. In the right hand figure, we could not account for additional brightness gradients from the BGG and very bright foreground star.}
    \label{fig:uncertain}
  \end{figure}
We obtain the following least-square fit for the colour magnitude relation (CMR) of the RS in the NGC 6482 group
  \begin{equation}
    B-R=(-0.029\pm0.008)\cdot M_R+(1.05\pm0.13)
  \end{equation}
with an RMS of 0.06\,mag. This fitted relation is represented by the solid black line in Fig.\,\ref{fig:cmd}. The index $r_{50}$ given to the colour legend ($B-R$) on the y-axis indicates that the values are the average within the half-light radius. The CMR fits well the relation defined by the brightest early-type galaxies in our sample, down to $M_R \sim -14$ mag and is also consistent with the colour distribution of the fainter galaxies - even though these exhibit a larger scatter. In the following, every object within the $3\sigma$ RMS of the fit and brighter than $M_R=-10.5\,$mag is considered to be a RS member of the group. Those galaxies are displayed as filled red data points in the CMD. We furthermore include two galaxies into this fiducial sample with uncertain photometry that formally place them redwards of the above $3\sigma$ range, but which morphologically resemble diffuse dwarf galaxies at the group's distance\,\footnote{These galaxies are listed in Tab.\,\ref{tab:appendix2} with IDs 20 ($M_R=14.2$\,mag, $B-R=2.48^{+0.12}_{-1.13}$\,mag) and 22 ($M_R=14.1$\,mag, $B-R=1.74\pm0.13$\,mag} (see also Fig.\,\ref{fig:uncertain}). Galaxies belonging to the blue cloud are labeled with blue datapoints. After selection around the red sequence, twelve out of 80 galaxies were considered to be background galaxies on the basis of their extreme red colours. Three of those are visible as black datapoints in the CMD. When only considering galaxies brighter than the 50\% completeness level at $M_R=-10.5$\,mag, 22 further galaxies are disregarded. Finally, including the two probable members with uncertain photometry mentioned above, this yields a fiducial sample of 48 probable group member galaxies (see Tab.\,\ref{tab:appendix2}).\\

\subsection{Surface brightness profiles measurements}\label{sec:sb}

Surface brightness (SB) profiles of all investigated galaxies were analyzed using the analytic expression suggested by \citet{sersic_1968}. In our case we fitted single S\'{e}rsic profiles to the $R$-band SB obtained from the undegraded images (see Sect.\,\ref{sec:photometry}), that is
\begin{equation}
\mu(R^c)=\mu_e+1.0857\cdot b_n\left[\left(\frac{R^c}{r_e}\right)^{1/n}-1\right].
\end{equation}
$\mu_e$ is the SB of the isophote at the effective radius $r_e$. The constant $b_n$ is defined in terms of the parameter $n$ which describes the shape of the light profile. As shown by \citet{caon_1993}, a convenient approximation
relating $b_n$ to the shape parameter $n$ is $b_n=1.9992n-0.3271$ for $1\lesssim n\lesssim10$, which we applied in our calculations. The $c$ in the variable $R^c$ shall denote that we performed the S\'{e}rsic fits with respect to the circularized radius $R^c=a\sqrt{1-\epsilon}$, where $a$ is the major axis of the isophote with its ellipticity $\epsilon$. We note that a multiple S\'{e}rsic fit would be more appropriate in the case of the S(lens)0 galaxy in our sample (ID 6 in Tab. \ref{tab:appendix2}) (see \citealt{kormendy_2012} and \citealt{janz_2012}). Thus, the errors of a single S\'ersic fit are rather large for this galaxy. For consistency with the rest of the sample we stick to a single fit as reference.
We note at this point the importance of a reliable background estimation for the SB fits to dwarf galaxies. \citet{caon_2005} show that incorrect sky background estimates lead to significant differences in the S\'{e}rsic fitting parameters. We are confident that the individual curve-of-growth-method described above gives a robust background estimate for each galaxy.\\
The S\'ersic fits were performed with different fitting ranges. The standard fit excluded the inner two arcseconds (i.e., one arcsecond of $R^c$), which is $\sim$3 times the seeing, and was performed until the intensity reaches $\mu_R=26.5$\,mag/arcsec$^2$. For some very faint SB profiles the fit did not converge so that either only the inner arcsecond was excluded or the limiting SB was set to $\mu_R=25.0$\,mag/arcsec$^2$. Nonetheless, we tried to perform the fit with all of the three settings to get robust error estimates that are provided in Tab.\,\ref{tab:appendix2}. For nucleated galaxies only the main body of the galaxy was fitted, excluding the central luminosity spike. Total luminosities were not computed from those fits, but from information of the total galaxy flux given by the {\verb*#ellipse#} outputs.\\
The results of these measurement are used to analyze the photometric scaling relations of dwarf galaxies in the fossil group NGC 6482, and to identify galaxies in our sample which exhibit centrally concentrated light profiles -- typical for intrisically luminous galaxies -- resulting in high S\'{e}rsic $n$ (larger than 2). In particular for galaxies with faint total magnitudes, this would indicate that they are background galaxies following \citet{lieder_2012}.

\section{Results}

The fact that galaxies are brighter in the $R$-band than in the $B$-band
($\sim1.5$ mag) roughly compensates the missing depth in the $R$-band
($\sim1.2\,$mag). Since the seeing is better in the $R$-band, we present
all results related to the according $R$-passband
quantities.

\subsection{Spatial distribution}
  \begin{figure}
   \centering
   \includegraphics[width=0.5\textwidth]{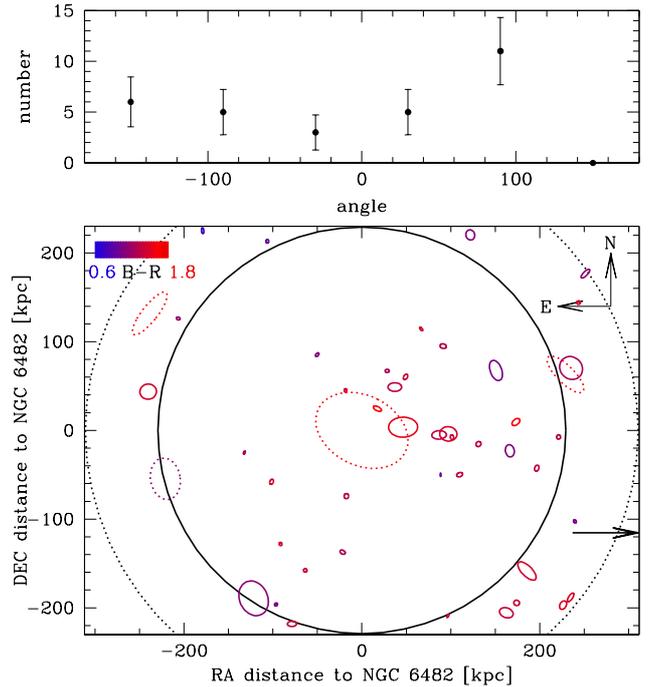}
      \caption{Lower panel: Spatial distribution of all galaxies considered group members (blue and red sequence, see Sect. \ref{sec:cmd}). Spectroscopically confirmed group members are denoted by dotted shapes. All objects are colour coded with respect to their $B-R$ colour and the size of each galaxy is scaled to its luminosity by roughly $L^{0.03}$. Ellipticity and position angle are represented by its value at the half-light radius. The solid circle represents the area which is completely covered by our field of view ($r=229$\,kpc). The size of the dotted circle corresponds to the virial radius ($r_{vir}=310$\,kpc). The arrow in the south-west corner indicates that another spectroscopically confirmed cluster member lies outside the field of view. Upper panel: Angular distribution of all galaxies shown in the lower panel within the solid circle. The 0 deg position is north in the lower panel and the angle grows clockwise. The errors in this plot are poissonian.}
         \label{fig:coord}
   \end{figure}   
In Fig.\,\ref{fig:coord} we present a $B-R$ colour coded, luminosity scaled spatial distribution of all galaxies within the field of view that are considered group members. Note that there is another spectroscopically confirmed group member outside the field of view with an apparent $B$-band magnitude of $m_B=15.5$\,mag \citep{zwicky_1963} -- comparable to the other spirals in this study.\\
The investigation of the angular distribution of the group galaxies (only within the fully covered circle of 229\,kpc) shows a preferred location of galaxies towards the west (90 deg position; see upper panel of Fig.\,\ref{fig:coord}). There are five galaxies in the far southwest end whose projection looks very clustered. Those five galaxies may constitute an intruding sub-group.\\
  \begin{figure}
    \centering
    \includegraphics[width=0.5\textwidth]{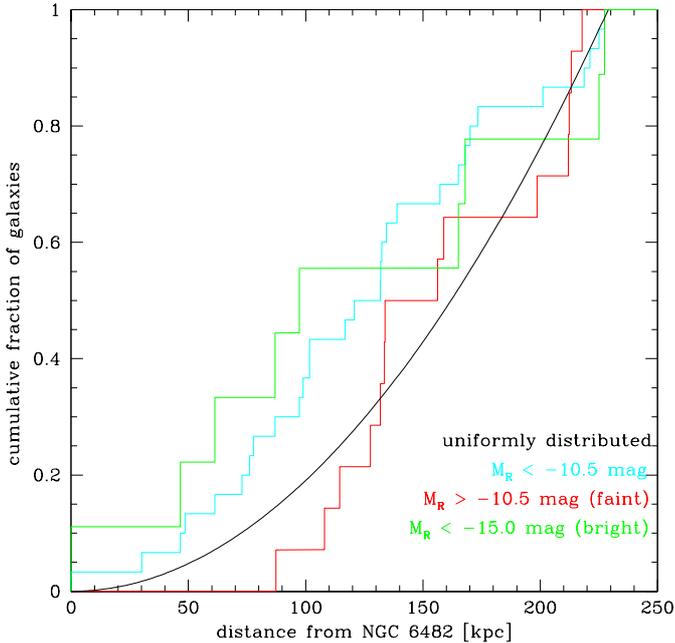}
    \caption{Radial distribution of investigated galaxies represented by their cumulative fraction with respect to their distance to the BGG. Only galaxies within the solid circle in Fig.\,\ref{fig:coord} (largest completely covered annulus) are taken into account as long as they obey the membership constrains of Sect. \ref{sec:cmd}. Green: bright galaxies ($M_R<-15\,$mag). Red: faint galaxies ($M_R>-10.5\,$mag). Light blue: all galaxies brighter than $M_R=-10.5$\,mag. The black solid line represents a uniform distribution as it would be the case for background galaxies.}
    \label{fig:cumrad}
  \end{figure}   
Another constraint on the group membership of the mentioned galaxies arises from Fig.\,\ref{fig:cumrad}. There we plot the cumulative radial distribution of galaxies of certain magnitude intervals with respect to their distance from the BGG, to test whether they are clustered towards the BGG. From this plot it is evident that the bright galaxies ($M_R<-15\,$mag, green histogram) are concentrated around the BGG as compared to an uniform distribution of galaxies which should follow the black line. This holds also for all galaxies we suppose to be group members (light blue histogram).

\subsection{Photometric scaling relations}
   \begin{figure}
   \centering
   \includegraphics[width=0.5\textwidth]{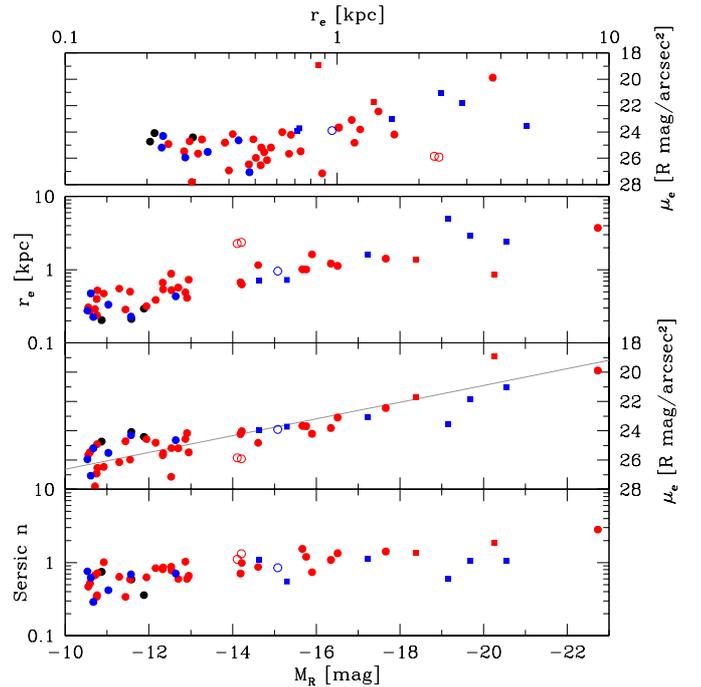}
      \caption{Photometric scaling relations of all investigated galaxies. Except the total magnitude (curve-of-growth), all quantities arise from single S\'ersic fits. In case of spiral galaxies, the fit was performed to the disk; for dE,N only the main body was fitted. Symbols as in Fig.\,\ref{fig:cmd}. While $r_e$ is the effective radius, $\mu_{e}$ represents the effective SB (at $r_e$). The gray line in the $\mu_e-M_R$-plot gives the relation of \citet{misgeld_2009}. The disrupted galaxy is not shown in this plot (no light profile available).}
         \label{fig:scaling}
   \end{figure}   
In Fig.\,\ref{fig:scaling} we present the most relevant photometric scaling relations of the galaxies presented in the CMD and the additional probable member with uncertain photometry at $B-R=2.48$, ID 20 in Tab.\,\ref{tab:appendix2}, including effective SB $\mu_e$ at the effective radius $r_e$, effective radius, S\'ersic index $n$, and total luminosity. The top panel in particular shows the correlation between effective SB and effective radius, also known as the Kormendy relation \citep{kormendy_1977}. We stress here that all quantities except the total magnitude are values which arise from single S\'ersic fits to the light profiles of the undegraded images (see Sect.\,\ref{sec:photometry}) down to $\mu_R=26.5$\,mag/arcsec$^2$. We disregard central bright components like nuclei, i.e., only the main body of the galaxy is considered. Galaxies whose properties constitute strong outliers in these plots are typically candidates for background galaxies. There is one source which is an outlier in three of the four plots: this is the S0 galaxy (MRK 0895), the host of the disrupted galaxy (see Sect. \ref{sec:disrupted}), a spectroscopically confirmed cluster member. It has a comparatively high surface brightness and small size compared to the main body of group member galaxies which might be related to the edge-on view, simply an effect of its high inclination (see Fig.\,\ref{fig:disrupted}).  We note that this galaxy is well fit by a single S\'{e}rsic profile (as seen in the right panel of Fig.\,\ref{fig:disrupted}). The fact that the grand design spiral galaxies tend to deviate from the photometric scaling relations given by the early-types is also visible in these plots. The spirals tend to have larger half-light radii and fainter effective SBs as expected from the early-type relations, resulting in lower concentration parameter $n$. This is not unexpected since our S\'ersic fits consider the disks only. Two other outliers represent the probable members with uncertain photometry denoted by open red circles. These are the galaxies displayed in Fig.\,\ref{fig:uncertain}. For the galaxy with two foreground stars in the center the low SB can be due to difficult masking which took most of its light.\\
Clear trends are visible in all relations in the sense that brighter galaxies are larger ($r_e$), have brighter SB at the half-light radius and more centrally concentrated light profiles (S\'{e}rsic $n$). There is no faint galaxy with high S\'{e}rsic $n$ which would qualify it as background galaxy.  This is a consequence of our applied $\mu-mag$ selection criterion (see Sect.\,\ref{sec:sample}) which initially rejects faint galaxies with high central light concentration. Another point is worth mentioning. We do not see any galaxy within the interval $-14<M_R<-13\,$mag. A similar dip in the galaxy luminosity function around this magnitude was reported by \citet{hilker_2003} for the Fornax cluster. We however refrain from addressing the statistical significance of the gap and its origin due to the low number statistics at these low luminosities.

  \subsection{Luminosity Function}
   \begin{figure}
   \centering
   \includegraphics[width=0.5\textwidth]{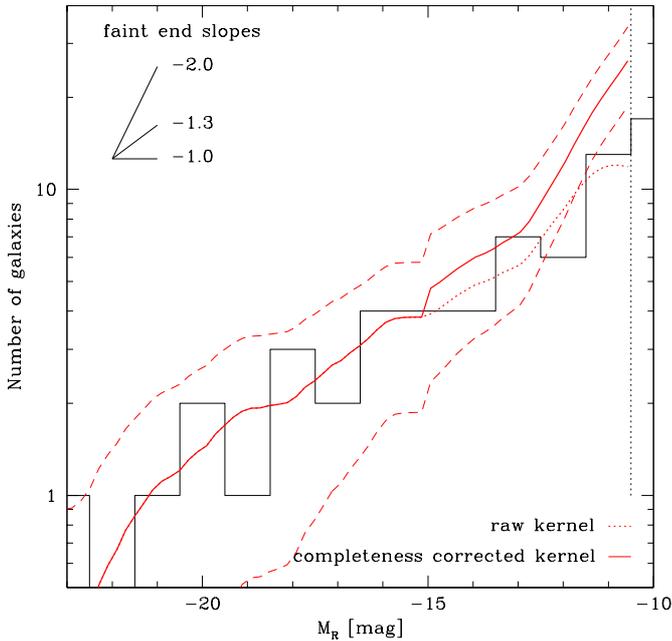}
      \caption{$R$-band luminosity function of member considered galaxies (see Fig.\,\ref{fig:cmd}). The black histogram represents the observed data (bin width: 1\,mag). The red dotted line is a binning independent representation of the counts (Epane\v{c}nikov kernel of 1\,mag bin width) while the red solid line is its completeness corrected (see Sect. \ref{sec:sample}) counterpart with the $1\sigma$ uncertainty limits (dashed). The vertical dotted line is our 50\% completeness limit at $M_R=-10.5\,$mag. For comparison some faint end slopes are illustrated in the top left corner of the plot. The best fit slope to our data is $\alpha \sim -1.3$, see text.}
         \label{fig:lf}
   \end{figure}   
In Fig.\,\ref{fig:lf} we show our completeness uncorrected galaxy luminosity function with the black histogram (1\,mag bin width, steps of 1\,mag). The sample used for the LF are the 48 galaxies considered as likely members, see Sect. 5.2. For a better visualization of the LF we use a binning independent sampling of the completeness corrected LF, performed by an Epane\v{c}nikov kernel \citep{epanecnikov_1969} with a bin width of $1.0$\,mag -- displayed with red colours in Fig.\,\ref{fig:lf}.\\
Because of the missing $L^*$ galaxies in a FG, the bright end of the LF looks different from normal cluster LFs. Thus, a Schechter fit to the LF will only be poorly constrained at the bright end. Nevertheless, it is meaningful for the faint end. We performed a fit to the galaxy number count distribution (assuming poissonian errors), including completeness correction for galaxies fainter than $M_R=-15\,$. The fitting interval was chosen to end with $M_R=-10.5\,$mag (our 50\% photometrical completeness limit). We fit the number count distribution of all galaxies in our imaging survey, noting that this only 84\% of $r_{vir}$ is covered by our dataset. The faint end slope of an error weighted Schechter function fit reveals $\alpha=-1.32\pm0.05$. A similar fit to the luminosity function within the circular region fully covered by our data ($0.74\,r_{vir}$, see solid line in Fig.\,\ref{fig:coord}), yields a marginally steeper slope $\alpha=-1.49\pm0.13$.\\
There is a hint for an upturn in the LF fainter than $M_R=-12\,$mag, but completeness correction starts to play an important role in this magnitude range, such that we will not discuss it in more detail.

\subsection{The Brightest Group Galaxy NGC\,6482}
  \begin{figure}
    \centering
    \includegraphics[width=0.5\textwidth]{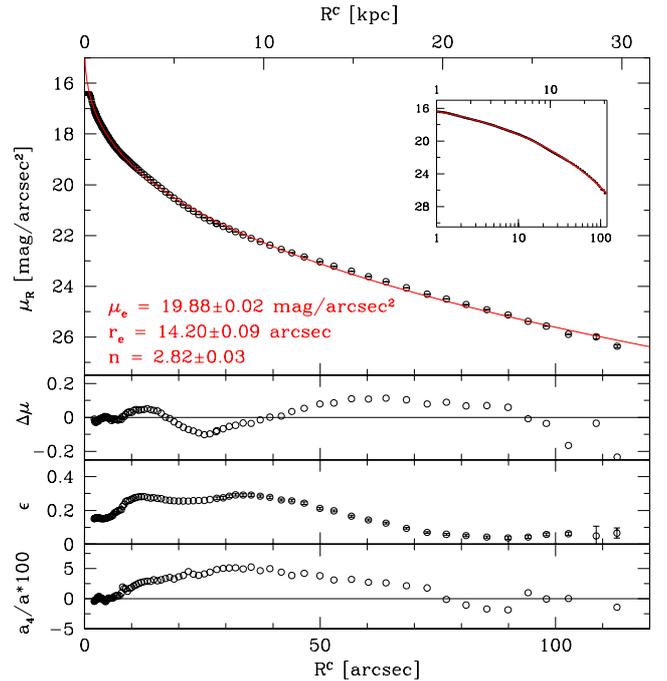}
    \caption{Radial profiles of the BGG NGC 6482. Top panel: SB profile, including a plot of logarithmic radial scale in the inset. In both cases the red line represents the single S\'{e}rsic fit, the fitting results are provided too. The second panel displays the deviation of the data from the S\'{e}rsic fit. The third panel shows the ellipticity with respect to radial distance and the bottom panel illustrates the $a_4/a$ parameter which describes the deviation of the isophote from a perfect ellipse.} 
    \label{fig:bgg}
  \end{figure}
The radial SB profile of the BGG (see Fig.\,\ref{fig:bgg}) is well described by a single S\'{e}rsic fit as the deviation of the photometric data to the fit (second panel) within the inner 100 arcsec remains at values smaller than $\Delta\mu=\pm0.1\,$mag/arcsec$^2$. We fit the light profile with respect to the circularized radius and exclude the inner 3 arcseconds (the inner 2 arcseconds are obviously affected by saturation) from the fit. Its S\'{e}rsic $n=2.82\pm0.03$ -- the light concentration parameter -- is rather small for such a giant elliptical galaxy, given that a S\'{e}rsic $n=4$ represents the typical de Vaucouleurs profile. In particular it is smaller than reported by \cite{alamo-martinez_2012} who find values of $\sim3.9$ in $g$- and $z$-band. However, smaller values of $n$ for fossil group central galaxies were already reported by the FOGO collaboration \citep{aguerri_2011,mendez-abreu_2012}, who obtain a mean S\'{e}rsic index of $n\sim3$ for a sample of 21 FGs.\\
The inner isophotes $(10''\lesssim R^c\lesssim50'')$ of NGC\,6482 exhibit a moderately large ellipticity $(\epsilon\sim0.3)$, and turn into almost spherical isophotes in the outskirts $(R^c>70'')$. The elevated overall ellipticity is accompanied by a disky shape in the inner galaxy part (lower panel in Fig.\,\ref{fig:bgg}). This is seen from the $a_4/a$ parameter, for which positive values denote disky isophotes \citep{bender_1988}.

\subsection{A disrupted galaxy around MRK\, 895}\label{sec:disrupted}
  \begin{figure*}
   \subfigure{\includegraphics[width=0.33\textwidth]{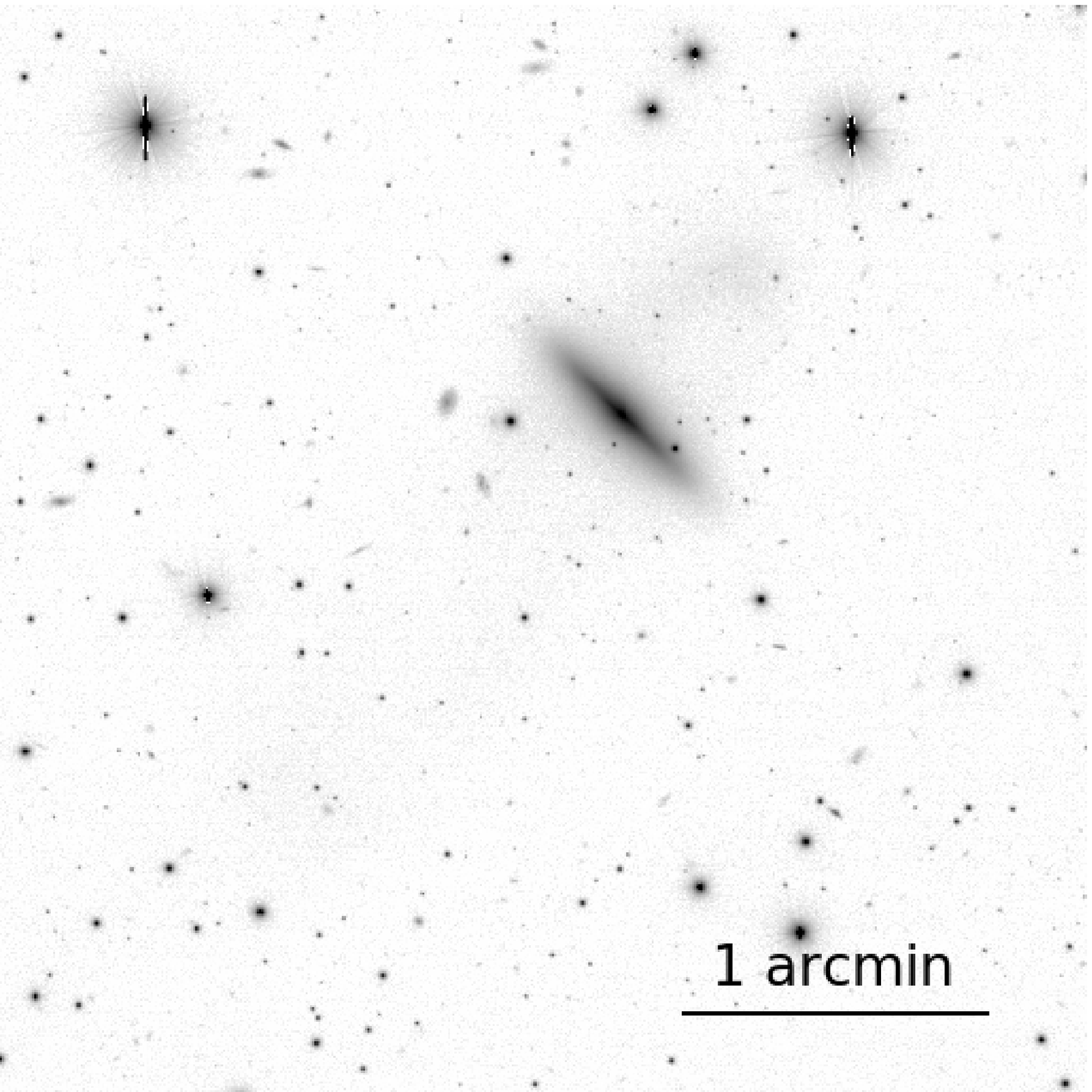}}\hfill
   \subfigure{\includegraphics[width=0.33\textwidth]{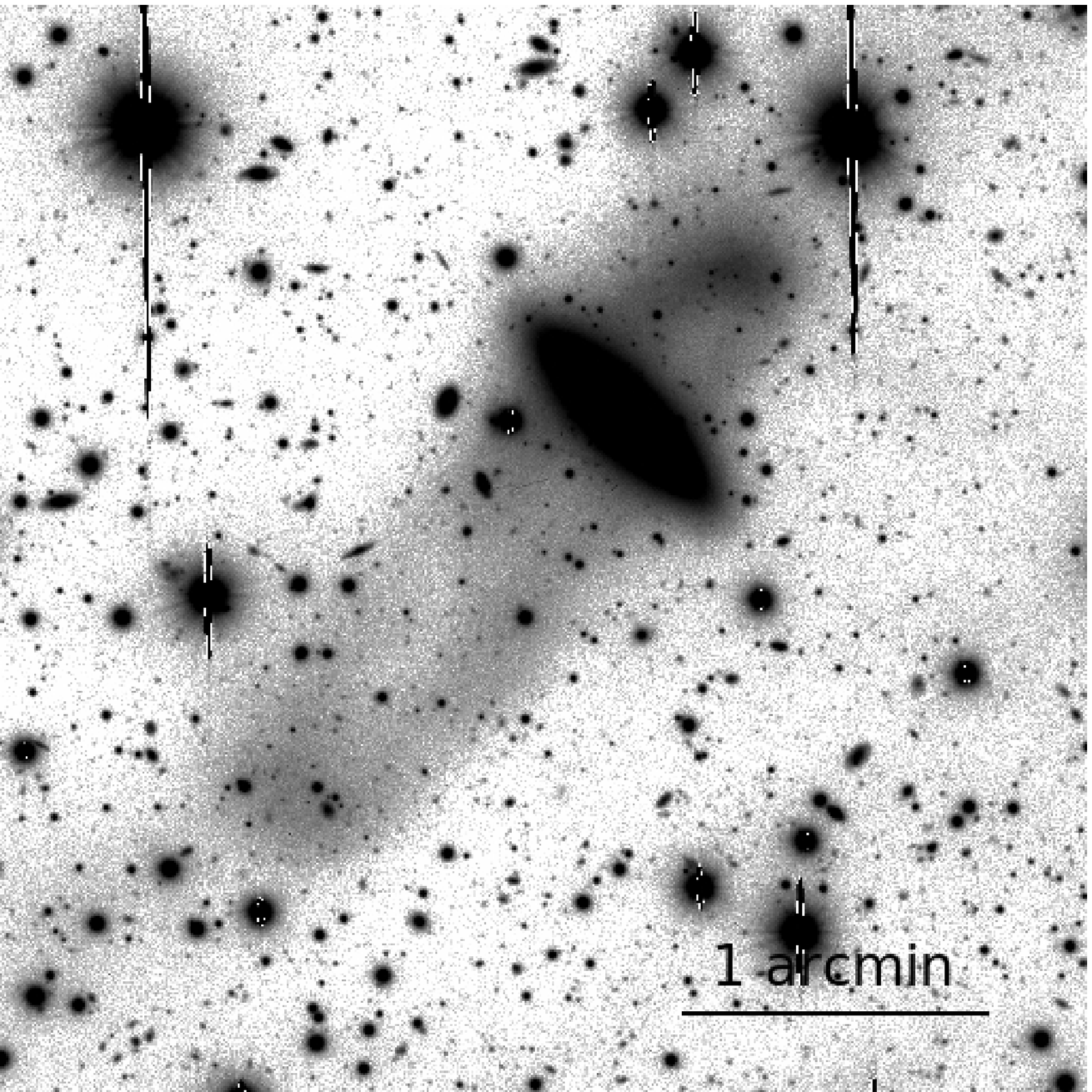}}\hfill
   \subfigure{\includegraphics[width=0.33\textwidth]{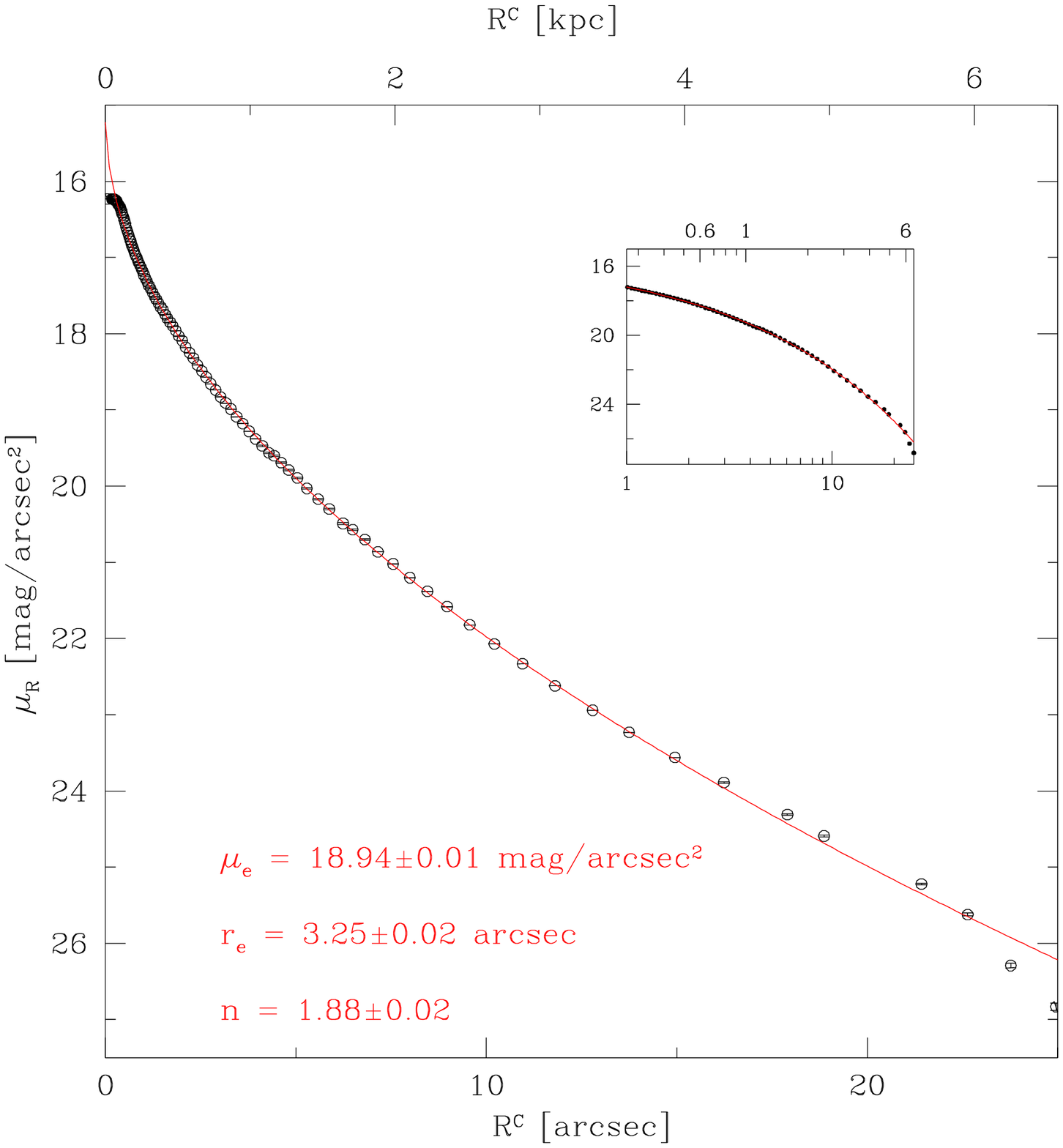}}\hfill
  \caption{Detection of a galaxy being disrupted. Left: S0 galaxy MRK\,895 by which the dwarf galaxy is being disrupted (North up, East left). Center: Image with different contrast settings to illustrate the tidal debris. Right: Light profile of the S0 galaxy. The red line visualizes the best S\'ersic model to the data.}
    \label{fig:disrupted}
  \end{figure*}

In one particular case we observe a galaxy being disrupted by the confirmed group member galaxy MRK\,895. This is an edge-on S0 galaxy as shown in Fig.\,\ref{fig:disrupted}. For the disrupted galaxy we can only provide SExtractor based photometry. We tuned SExtractor to detect the whole debris. The obtained flux represents only a lower limit because we have masked point sources as well as the hosting S0 galaxy. This fact is denoted by the arrow on this galaxy in the colour-magnitude diagram (Fig.\,\ref{fig:cmd}).\\
We masked the tidal debris for the \verb*#ellipse# investigation of the S0 galaxy in the outer regions, but did not mask it where the debris crosses the bright isophotes along the disk of the galaxy. Thus, the isophotes might be affected by light from the tidal debris. But this effect should not be significant for the total brightness of the S0 galaxy since the galaxy itself is very bright (see light profile in the right panel of Fig.\,\ref{fig:disrupted}).\\
The projected diameter of the debris is $\sim 35\,$kpc. We investigated the $B-R$ colours of both the debris and the overdensity close to the North-West corner of MRK 0895. While the overdensity shows $B-R\sim1.4\,$mag, the rest of the debris shows $B-R$ colours of $\sim1.2\,$ mag. The S0 galaxy itself has $B-R\sim1.69\,$ mag as shown in Fig.\,\ref{fig:cmd} (The open blue square represents the disrupted galaxy and the third brightest galaxy is the host). Hence, the debris is on average $0.3\,$mag bluer than its host. The disrupted galaxy is obviously not on the red sequence. The progenitor could have been a relatively luminous late-type and/or metal-poor system, since the flux of the whole debris adds to $M_R=-17.7\,$mag.

\section{Discussion and Conclusions}
\subsection{Brightest Group Galaxy}
We find $B-R=1.73$ for the BGG NGC 6482, a colour typical for metal-rich old stellar populations. Such a red colour was also found for NGC 6482 by \cite{alamo-martinez_2012} ($g-z=1.85$ at $r_{50}$). Another finding of our study is the relatively low S\'ersic index of $n\sim2.82$ for NGC 6482, in agreement with the studies of the FOGO group who find S\'ersic indices around $n\sim3$ for FGs \citep{aguerri_2011,mendez-abreu_2012}. They argue based on \citet{hopkins_2009} and \citet{kormendy_2009} that low S\'ersic indices for giant ellipticals ($n\sim2.5$) are tracers for dissipational mergers. 
\citet{bender_1988} and \citet{khochfar_2005} find that disky isophotes of giant elliptical galaxies are the result of wet, gas-rich mergers, i.e., mergers with participation of spiral or irregular galaxies. We clearly see disky isophotes in the BGG ($a_4/a\sim0.05$) and \citet{alamo-martinez_2012} identify a dust lane in their HST images of NGC\,6482 -- favoring the merger scenario of our FG as originally claimed by \citet{ponman_1994}. But that does not necessarily disfavor the "failed group" scenario of \citet{mulchaey_1999}. \cite{dekel_2009} suggested that inflows of cold streams might have been the major formation scenario in the early Universe. A rotational disk would have kept intact while giant star-forming clumps merge into the center to form a massive spheroid. These star-forming clumps -- one could call them galaxies -- could leave an imprint on the BGG's morphology like disky isophotes or dust lanes. \cite{oser_2010} show in simulations that the majority of stars of an intermediate mass central galaxy (like our NGC\,6482) have been formed ex-situ in clumps, thus they have been accreted. In this sense, a massive dark matter halo accompanied by cold gas streams would reflect the "failed group" scenario but should also show merger signatures. A difference in both formation scenarios would then become washed-out.

\subsection{Photometric scaling relations}
We adopt the $\mu$-mag relation of \citet{misgeld_2009} to simulate galaxies in order to discover dwarf galaxies in the NGC\,6482 group. These relations agree well with the sample properties of the recovered dwarfs as seen in Fig.\,\ref{fig:scaling}. Disregarding disky galaxies (i.e., spirals and S0s), most galaxies in the sample follow the same photometric scaling relations, indicating that they truly belong to the NGC\,6482 group. However, at the faint end of the distribution background contamination should play an important role. We accounted for that by applying a certain colour range for group members, following the CMR of the bright galaxies.\\
In general, the photometric scaling relations are very similar to those found in galaxy clusters. This includes the distribution in CM space (see Fig.\,\ref{fig:cmd}) where both the red and blue sequence are distinctly defined. Our best fit of the CMR to the RS has a slope of $-0.029\pm0.008$, comparable to similar slope determinations in $B-R/R$ in some galaxy clusters($-0.045 \pm 0.028$ for Coma; \citealp{adami_2006}; ($-0.055 \pm 0.009$ for Perseus; \citealp{conselice_2002}), possibly somewhat at the shallow side.

\subsection{Luminosity function}

\begin{table}
\caption{Luminosity function faint end slopes in literature.}             
\label{tab:slopes}      
  \centering                          
  \begin{tabular}{l c c l}        
    \hline\hline                 
    group/cluster & $\alpha$ & band &  reference\\    
    \hline                        
    Virgo & $-1.30$ & $B$ &\cite{sandage_1985} \\
    Coma & $-1.41$ & $R$ & \cite{secker_1997} \\
    Hydra I& $-1.13$ & $V$ & \cite{misgeld_2008} \\
    Centaurus & $-1.14$ & $V$ & \cite{misgeld_2009} \\
    Hercules & $-1.29$ & $V$ & \cite{sanchez-janssen_2005}\\
    Fornax & $-1.1$ & $V$ &\cite{hilker_2003}\\
    Perseus &  $-1.26$ & $B$ & \cite{penny_2008}\\
    \hline
    diff. env.$^\dagger$ & $-1.19$ & $R$ & \cite{trentham_2002}\\
    \hline 
    2dF & $-1.28$ & $b_J$ & \cite{de_propris_2003}\\                                      
    \hline
    NGC\,6482 & $-1.32$ & $R$ & this study\\ 
    \hline 

  \end{tabular}
  \tablefoot{Faint end slopes $\alpha$ were derived by a single Schechter function fit. $^\dagger$ Composite LF (linear fit) of five different environments with varying galaxy density.}
\end{table}

Little is known about the faint satellite systems of FGs. The deepest observational study of a FG in the literature reaches $M_g=-16\,$mag \citep{mendes_de_oliveira_2006} -- that is four magnitudes fainter than the group's characteristic magnitude ($M^*+4\,$mag). Therefore no solid constraints on the dwarf galaxy regime can be drawn from these studies. Particularly the faint end slope of the LF has been poorly constrained up to now by other FG studies, and varies strongly from $-0.6$ to $-1.6$ \citep{cypriano_2006,mendes_de_oliveira_2006,mendes_de_oliveira_2009,proctor_2011,eigenthaler_2012}.\\
Our study is, to our knowledge, the first one to probe deep into the dwarf galaxy regime of a fossil group, extending the available literature studies by 6-7 mag in total luminosity. Our investigation reveals dwarf galaxies as faint as $M_R=-10.5\,$mag. We find a faint end slope of $\alpha=-1.32 \pm 0.05$, fully within the range of values typically found in cluster environments (Tab.\ref{tab:slopes}) like Coma or more unevolved clusters like Virgo and Hercules. The faint end slope of a composite LF, averaged over 60 nearby galaxy clusters, reveals a similar value of $\alpha=-1.28$ \citep{de_propris_2003}. Previous studies suggested that indeed the faint end slope of the LF is independent on environment: \cite{trentham_2002} show such an almost invariable faint end slope in their study of five different environments with varying galaxy density and morphological content. They find a composite faint end slope of $\alpha=-1.19$ for their entire sample. The result of our study, the first deep one in a fossil group, is consistent with the average slope found in a range of environments, and lends further credence to the notion that the faint end slope of the galaxy luminosity function depends only very little on environment. \\
Another useful parameter to describe the LF shape is the dwarf-to-giant ratio \citep{phillipps_1998, sanchez-janssen_2008}. We compare our data with the study of \cite{trentham_2002}, adopting their definition of the d/g ratio as $d/g=N(-17<M_R<-11)/N(M_R<-17)$. By taking into account the completeness correction we obtain: $d/g=4.1\pm0.6$. The error acounts for color and magnitude uncertainties which could propagate into our group membership determination. The dwarf-to-giant ratio we find is consistent with the average $d/g=3.2$ (rms 1.2) reported in \cite{trentham_2002}, and in particular matches the values they find for virialized systems like NGC\,1407 and the Virgo cluster.\\
We do not find any galaxy in the magnitude range of $-14<M_R<-13\,$mag, best seen in Fig.\,\ref{fig:scaling}. While this interesting feature of NGC\,6482 might be due to the low number counts (Fig.\,~\ref{fig:lf}), we note that a similar dip has been found by \cite{hilker_2003} in the Fornax cluster. As already noted by these authors: this magnitude range is the transition from dE to dSph.\\
Another two properties of dwarf galaxies are the fraction of early- to late-types and the fraction of nucleated dEs. \cite{trentham_2002} found evidence that dynamically more evolved systems have a higher fraction of dE as compared to dIrr (see also \citealt{mahdavi_2005}). The percentage of dwarfs in the range $-17<M_R<-11\,$mag classified dE as opposed to dIrr is $89^{+1}_{-3}$ percent, comparable to the fraction \cite{trentham_2002} find in the central 200\,kpc of Virgo cluster. The errorbars arise from photometric errors. Among the dEs $38\%$ are nucleated within the same magnitude range. This is comparable to the 40\% of a combined nucleation rate of four groups in the \cite{trentham_2002} sample but only half the nucleation they find for the Virgo cluster (70\%). \\\\
\noindent We conclude that the NGC\,6482 fossil group shows photometric properties consistent with those of regular galaxy clusters and groups, including a normal abundance of faint satellites. The potential `missing satellite problem' in this fossil group is thus of similar scale than in other environments.

\begin{acknowledgements}
Special thanks are addressed to Mischa Schirmer for his support on THELI issues with the tricky Subaru data. We also wish to thank the referee, Jairo M\'endez-Abreu, for useful comments to improve the quality of this manuscript.\\
S.L.\ is supported by the ESO Studentship Program. S.L.\ and T.L.\ are supported within the framework of the Excellence Initiative by the German Research Foundation (DFG) through the Heidelberg Graduate School of Fundamental Physics (grant number GSC 129/1). This research has made use of the NASA/IPAC Extragalactic Database (NED) which is operated by the Jet Propulsion Laboratory, California Institute of Technology, under contract with the National Aeronautics and Space Administration. 
\end{acknowledgements}

\bibliographystyle{aa} 

\Online
\begin{appendix}
\section{Tables}
\onllongtab{5}{
\begin{longtable}{cccccccccc}
\caption{Properties of all galaxies considered as group members in this study}\\
\multicolumn{10}{l}{ID: galaxy identification number in our catalogue sorted by $M_R$}\\
\multicolumn{10}{l}{$\alpha$ (J2000): Right Ascension}\\
\multicolumn{10}{l}{$\delta$ (J2000): Declination}\\
\multicolumn{10}{l}{$M_R$: absolute $R$-band magnitude (adopted $m-M=33.71$ mag), errorbars do not include photometric calibration uncertainty of $\sigma_R=0.04\,$mag.}\\
\multicolumn{10}{l}{$M_B$: absolute $B$-band magnitude (adopted $m-M=33.71$ mag); errorbars do not include photometric calibration uncertainty of $\sigma_B=0.12\,$mag.}\\
\multicolumn{10}{l}{$B-R$: integrated $B-R$ colour (within half-light radius $r_{50}$)}\\
\multicolumn{10}{l}{$\mu_{e}$: effective surface brightness in $R$-band from S\'ersic fit in $R$-band applied to circularized isophotes}\\
\multicolumn{10}{l}{$r_{e}$: effective radius from S\'ersic fit in $R$-band applied to circularized isophotes, errors include uncertainty introduced by seeing.}\\
\multicolumn{10}{l}{$n$: S\'ersic index}\\
\multicolumn{10}{l}{Type: classification type (for member considered galaxies only; all early-type dwarf galaxies are labeled dE)}\vspace{0.5cm}\\

\hline
\hline
ID & $\alpha$ (J2000) & $\delta$ (J2000) & $M_R$ & $M_B$ & $B-R$ & $\mu_{e}$ & $r_{e}$ & n & Type\\
 & [deg] & [deg] & [mag] & [mag]& [mag] & [mag/arcsec$^2$] & [kpc] & & \\ 
\hline\\
\endfirsthead
\caption{Continued.} \\
\hline
ID & $\alpha$ (J2000) & $\delta$ (J2000) & $M_R$ & $M_B$ & $B-R$ & $\mu_{e}$ & $r_{e}$ & n & Type\\
 & [deg] & [deg] & [mag] & [mag]& [mag] & [mag/arcsec$^2$] & [kpc] & & \\ 
\hline\\
\endhead
\hline
\endfoot
\endlastfoot 
 1$^*$ & 267.95346 & 23.07192 & -22.73 $\pm$ 0.02 & -21.09 $\pm$ 0.02 & 1.73 $\pm$ 0.02$^{\ddagger}$ & 19.88 $\pm$ 0.02 &  3.73 $\pm$ 0.13 & 2.82 $\pm$ 0.03 	& E\\
 2$^*$ & 268.22717 & 23.21117 & -20.54 $\pm$ 0.04 & -18.92 $\pm$ 0.07 & 1.72 $\pm$ 0.07 & 21.01 $\pm$ 0.01 &  2.42 $\pm$ 0.13 & 1.05 $\pm$ 0.01 	& Sc\\
 3$^*$ & 267.69073 & 23.13864 & -20.26 $\pm$ 0.02 & -18.64 $\pm$ 0.03 & 1.69 $\pm$ 0.02 & 18.94 $\pm$ 0.01 &  0.85 $\pm$ 0.13 & 1.88 $\pm$ 0.02 	& S0\\
 4$^*$ & 268.20694 & 23.01428 & -19.69 $\pm$ 0.04 & -18.53 $\pm$ 0.05 & 1.19 $\pm$ 0.06 & 21.84 $\pm$ 0.02 &  2.88 $\pm$ 0.13 & 1.07 $\pm$ 0.03 	& SBc\\
 5 & 268.09314 & 22.87197 & -19.15 $\pm$ 0.09 & -17.97 $\pm$ 0.09 & 1.19 $\pm$ 0.07 & 23.51 $\pm$ 0.03 &  4.99 $\pm$ 0.14 & 0.60 $\pm$ 0.03 	& SBc\\
 6 & 267.90015 & 23.07572 & -18.38 $\pm$ 0.01 & -16.75 $\pm$ 0.03 & 1.60 $\pm$ 0.01 & 21.71 $\pm$ 0.02 &  1.36 $\pm$ 0.13 & 1.38 $\pm$ 0.03 	& S(lens)0\\
 7 & 267.68355 & 23.14623 & -17.70$^{-0.5}_{+0.01}$ & -16.37 $^{-0.5}_{+0.02}$ & 1.33 $\pm$ 0.13$^{\dagger}$ & --- &  --- & --- & disrupted\\
 8 & 267.74030 & 22.90452 & -17.66 $\pm$ 0.02 & -16.25 $\pm$ 0.02 & 1.53 $\pm$ 0.02 & 22.44 $\pm$ 0.01 &  1.42 $\pm$ 0.13 & 1.42 $\pm$ 0.01 	& dE,N\\
 9 & 267.78040 & 23.14328 & -17.23 $\pm$ 0.05 & -16.17 $\pm$ 0.05 & 1.07 $\pm$ 0.05 & 23.05 $\pm$ 0.01 &  1.59 $\pm$ 0.13 & 1.12 $\pm$ 0.02 	& dIrr\\
 10 & 267.84186 & 23.06776 & -16.51 $\pm$ 0.02 & -15.08 $\pm$ 0.04 & 1.55 $\pm$ 0.01 & 23.09 $\pm$ 0.01 &  1.13 $\pm$ 0.13 & 1.34 $\pm$ 0.01 	& dE,N\\
 11 & 268.22913 & 23.11805 & -16.35 $\pm$ 0.03 & -14.96 $\pm$ 0.06 & 1.52 $\pm$ 0.02 & 23.81 $\pm$ 0.01 &  1.22 $\pm$ 0.13 & 1.09 $\pm$ 0.01 	& dE,N\\
 12 & 267.85382 & 23.06654 & -15.90 $\pm$ 0.04 & -14.53 $\pm$ 0.05 & 1.42 $\pm$ 0.03 & 24.20 $\pm$ 0.01 &  1.62 $\pm$ 0.13 & 0.74 $\pm$ 0.01 	& dE\\
 13 & 267.76700 & 22.85475 & -15.76 $\pm$ 0.05 & -14.38 $\pm$ 0.08 & 1.47 $\pm$ 0.02 & 23.68 $\pm$ 0.01 &  1.01 $\pm$ 0.13 & 1.20 $\pm$ 0.02 	& dE,N\\
 14 & 267.91089 & 23.12349 & -15.67 $\pm$ 0.04 & -14.35 $\pm$ 0.06 & 1.41 $\pm$ 0.03 & 23.66 $\pm$ 0.02 &  1.02 $\pm$ 0.13 & 1.54 $\pm$ 0.02 	& dE,N\\
 15 & 267.76242 & 23.04759 & -15.30 $\pm$ 0.05 & -14.17 $\pm$ 0.06 & 1.11 $\pm$ 0.08 & 23.72 $\pm$ 0.02 &  0.73 $\pm$ 0.13 & 0.55 $\pm$ 0.01 	& dIrr\\
 16 & 267.66498 & 23.25852 & -15.08 $\pm$ 0.07 & -13.90 $\pm$ 0.08 & 1.19 $\pm$ 0.03 & 23.90 $\pm$ 0.02 &  0.96 $\pm$ 0.13 & 0.85 $\pm$ 0.02 	& dIrr?\\
 17 & 267.81354 & 23.30462 & -14.64 $\pm$ 0.03 & -13.48 $\pm$ 0.05 & 1.14 $\pm$ 0.03 & 23.95 $\pm$ 0.02 &  0.71 $\pm$ 0.13 & 1.08 $\pm$ 0.04 	& dIrr\\
 18 & 267.68402 & 22.87308 & -14.61 $\pm$ 0.07 & -13.09 $\pm$ 0.09 & 1.55 $\pm$ 0.02 & 24.82 $\pm$ 0.02 &  1.16 $\pm$ 0.13 & 0.87 $\pm$ 0.01 	& dE\\
 19 & 268.04355 & 22.84158 & -14.22 $\pm$ 0.05 & -12.83 $\pm$ 0.12 & 1.48 $\pm$ 0.04 & 24.01 $\pm$ 0.01 &  0.63 $\pm$ 0.13 & 0.99 $\pm$ 0.01 	& dE,N\\
 20 & 267.75461 & 23.08187 & -14.21 $\pm$ 0.08 & -12.92 $\pm$ 0.25 & 2.48$^{+0.12}_{-1.13}$ & 25.91 $\pm$ 0.09 &  2.37 $\pm$ 0.20 & 1.32 $\pm$ 0.06 	& dE\\
 21 & 267.69376 & 22.86386 & -14.19 $\pm$ 0.07 & -12.70 $\pm$ 0.08 & 1.57 $\pm$ 0.03 & 24.22 $\pm$ 0.01 &  0.68 $\pm$ 0.13 & 0.71 $\pm$ 0.01 	& dE,N\\
 22 & 267.93338 & 23.09780 & -14.11 $\pm$ 0.15 & -12.52 $\pm$ 0.35 & 1.74 $\pm$ 0.13 & 25.84 $\pm$ 0.14 &  2.28 $\pm$ 0.27 & 1.11 $\pm$ 0.09 	& dE,N\\
 23 & 267.84854 & 23.17222 & -12.95 $\pm$ 0.21 & -11.61 $\pm$ 0.19 & 1.38 $\pm$ 0.09 & 25.47 $\pm$ 0.02 &  0.73 $\pm$ 0.13 & 0.66 $\pm$ 0.02 	& dE\\
 24 & 267.72772 & 23.02693 & -12.91 $\pm$ 0.10 & -11.60 $\pm$ 0.14 & 1.41 $\pm$ 0.05 & 24.16 $\pm$ 0.01 &  0.41 $\pm$ 0.13 & 0.60 $\pm$ 0.02 	& dE\\
 25 & 267.82730 & 23.01918 & -12.87 $\pm$ 0.05 & -11.69 $\pm$ 0.10 & 1.43 $\pm$ 0.04 & 24.56 $\pm$ 0.02 &  0.49 $\pm$ 0.13 & 1.03 $\pm$ 0.02 	& dE\\
 26 & 267.89713 & 23.13568 & -12.70 $\pm$ 0.08 & -11.37 $\pm$ 0.15 & 1.48 $\pm$ 0.07 & 25.19 $\pm$ 0.02 &  0.57 $\pm$ 0.13 & 0.60 $\pm$ 0.02 	& dE,N\\
 27 & 268.15881 & 23.30970 & -12.64 $\pm$ 0.09 & -11.73 $\pm$ 0.12 & 0.78 $\pm$ 0.07 & 24.64 $\pm$ 0.01 &  0.43 $\pm$ 0.13 & 0.71 $\pm$ 0.02 	& dE\\
 28 & 267.75369 & 22.86655 & -12.54 $\pm$ 0.06 & -11.17 $\pm$ 0.13 & 1.50 $\pm$ 0.04 & 25.18 $\pm$ 0.02 &  0.53 $\pm$ 0.13 & 0.78 $\pm$ 0.02 	& dE\\
 29 & 267.97821 & 22.92692 & -12.53 $\pm$ 0.18 & -11.21 $\pm$ 0.32 & 1.31 $\pm$ 0.11 & 27.14 $\pm$ 0.08 &  0.88 $\pm$ 0.14 & 0.88 $\pm$ 0.06 	& dE\\
 30 & 267.80301 & 23.05564 & -12.34 $\pm$ 0.07 & -11.12 $\pm$ 0.22 & 1.43 $\pm$ 0.05 & 25.53 $\pm$ 0.04 &  0.54 $\pm$ 0.13 & 0.86 $\pm$ 0.04 	& dE,N\\
 31 & 268.06989 & 23.01075 & -12.33 $\pm$ 0.06 & -10.94 $\pm$ 0.17 & 1.52 $\pm$ 0.06 & 25.66 $\pm$ 0.04 &  0.67 $\pm$ 0.13 & 0.81 $\pm$ 0.03 	& dE,N\\
 32 & 267.97345 & 22.99362 & -12.16 $\pm$ 0.04 & -10.83 $\pm$ 0.10 & 1.37 $\pm$ 0.03 & 24.81 $\pm$ 0.02 &  0.39 $\pm$ 0.13 & 0.84 $\pm$ 0.03 	& dE\\
 33 & 267.83710 & 23.06381 & -11.94 $\pm$ 0.04 & -10.64 $\pm$ 0.10 & 1.46 $\pm$ 0.03 & 24.57 $\pm$ 0.01 &  0.32 $\pm$ 0.13 & 0.63 $\pm$ 0.01 	& dE\\
 34 & 268.01102 & 23.16201 & -11.57 $\pm$ 0.09 & -10.63 $\pm$ 0.08 & 1.12 $\pm$ 0.05 & 24.30 $\pm$ 0.02 &  0.23 $\pm$ 0.13 & 0.69 $\pm$ 0.04 	& dE\\
 35 & 267.69958 & 23.06408 & -11.55 $\pm$ 0.08 & -10.36 $\pm$ 0.17 & 1.36 $\pm$ 0.08 & 25.97 $\pm$ 0.07 &  0.50 $\pm$ 0.13 & 0.59 $\pm$ 0.05 	& dE\\
 36 & 267.87686 & 23.19275 & -11.44 $\pm$ 0.07 & -10.06 $\pm$ 0.12 & 1.54 $\pm$ 0.11 & 24.71 $\pm$ 0.01 &  0.29 $\pm$ 0.13 & 0.34 $\pm$ 0.01 	& dE\\
 37 & 267.92081 & 23.14277 & -11.29 $\pm$ 0.09 & -10.20 $\pm$ 0.24 & 1.34 $\pm$ 0.07 & 26.14 $\pm$ 0.13 &  0.55 $\pm$ 0.14 & 0.64 $\pm$ 0.09 	& dE\\
 38 & 268.07532 & 23.29704 & -11.03 $\pm$ 0.14 & -10.21 $\pm$ 0.15 & 0.96 $\pm$ 0.12 & 25.51 $\pm$ 0.03 &  0.33 $\pm$ 0.13 & 0.42 $\pm$ 0.02 	& dE\\
 39 & 268.19034 & 23.20517 & -10.92 $\pm$ 0.12 & -9.80 $\pm$ 0.44 & 1.23 $\pm$ 0.10 & 26.46 $\pm$ 0.37 &  0.47 $\pm$ 0.17 & 1.01 $\pm$ 0.33 	& dE,N\\
 40 & 268.02652 & 22.90529 & -10.77 $\pm$ 0.18 & -9.20 $\pm$ 0.32 & 1.49 $\pm$ 0.15 & 26.53 $\pm$ 0.34 &  0.52 $\pm$ 0.20 & 0.72 $\pm$ 0.22 	& dE\\
 41 & 267.84235 & 22.85096 & -10.76 $\pm$ 0.09 & -9.47 $\pm$ 0.11 & 1.52 $\pm$ 0.10 & 24.92 $\pm$ 0.02 &  0.24 $\pm$ 0.13 & 0.36 $\pm$ 0.02 	& dE\\
 42 & 267.67389 & 23.22414 & -10.75 $\pm$ 0.14 & -8.78 $\pm$ 0.25 & 1.54 $\pm$ 0.18 & 26.92 $\pm$ 0.21 &  0.40 $\pm$ 0.16 & 0.34 $\pm$ 0.11 	& dE\\
 43 & 267.97455 & 23.11946 & -10.71 $\pm$ 0.14 & -9.29 $\pm$ 0.33 & 1.46 $\pm$ 0.21 & 27.79 $\pm$ 0.27 &  0.29 $\pm$ 0.15 & 0.67 $\pm$ 0.16 	& dE\\
 44 & 267.67850 & 22.96362 & -10.67 $\pm$ 0.09 & -9.76 $\pm$ 0.16 & 1.07 $\pm$ 0.07 & 25.19 $\pm$ 0.01 &  0.23 $\pm$ 0.13 & 0.29 $\pm$ 0.01 	& dE\\
 45 & 267.85178 & 23.01907 & -10.61 $\pm$ 0.17 & -9.83 $\pm$ 0.22 & 0.95 $\pm$ 0.12 & 27.06 $\pm$ 0.14 &  0.48 $\pm$ 0.14 & 0.62 $\pm$ 0.08 	& dE\\
 46 & 268.05835 & 22.93668 & -10.59 $\pm$ 0.10 & -9.36 $\pm$ 0.21 & 1.53 $\pm$ 0.09 & 25.47 $\pm$ 0.04 &  0.27 $\pm$ 0.13 & 0.52 $\pm$ 0.05 	& dE\\
 47 & 268.10507 & 23.04556 & -10.55 $\pm$ 0.20 & -9.52 $\pm$ 0.25 & 1.31 $\pm$ 0.12 & 25.64 $\pm$ 0.04 &  0.31 $\pm$ 0.13 & 0.47 $\pm$ 0.04 	& dE\\
 48 & 268.06409 & 22.86462 & -10.53 $\pm$ 0.30 & -9.84 $\pm$ 0.29 & 1.08 $\pm$ 0.16 & 25.94 $\pm$ 0.08 &  0.28 $\pm$ 0.13 & 0.76 $\pm$ 0.08 	& dE\\
\hline 
\vspace{0.1cm}\\
\multicolumn{10}{l}{{* objects with NED listed velocities, qualifing the galaxy as group member}}\\
\multicolumn{10}{l}{{$\dagger$ colour determined using SExtractor}}\\
\multicolumn{10}{l}{{$\ddagger$ colour determined \emph{at} half-light radius}}\\

\label{tab:appendix2}
\end{longtable}
}
\end{appendix}
\end{document}